\pdfoutput=1
\documentclass[runningheads,orivec]{llncs}
\usepackage{CJK}
\usepackage{listings}
\usepackage[table]{xcolor}
\usepackage{array}
\usepackage{pgf,tikz}
\usetikzlibrary{positioning, shapes, shapes.geometric, shapes.symbols, shapes.arrows, arrows.meta, decorations.pathreplacing, calc, fit, backgrounds, matrix, tikzmark, graphs, patterns}
\usepackage{float}
\usepackage{enumitem}

\usepackage{amsthm}
\usepackage{amssymb}
\usepackage{amsmath}
\usepackage{algorithm}
\usepackage[noend]{algpseudocode}
\usepackage{comment}
\usepackage{xparse}
\usepackage{xspace}

\usepackage{listings}
\usepackage{color}
\usepackage{mathtools}
\usepackage{multirow}
\usepackage{graphics}
\usepackage{paracol}
\usepackage{stmaryrd}
\usepackage{subcaption}
\usepackage{booktabs}
\usepackage{tabularx}
\usepackage{caption}
\usepackage[normalem]{ulem}
\usepackage{xurl}
\usepackage{hyperref}
\hypersetup{hypertexnames=false}
\usepackage{mathpartir}

\usepackage[capitalise]{cleveref}
\makeatletter
\lst@AddToHook{Init}{\def\@currentcounter{lstnumber}}
\makeatother
\lccode`\/=0

\definecolor{dkgreen}{rgb}{0,0.6,0}
\definecolor{gray}{rgb}{0.5,0.5,0.5}
\definecolor{mauve}{rgb}{0.58,0,0.82}
\definecolor{brightcerulean}{rgb}{0.11,0.67,0.84}
\definecolor{cerulean}{rgb}{0.0, 0.48, 0.65}

\lstdefinestyle{mystyle}{frame=none,
  aboveskip=3mm,
  belowskip=3mm,
  showstringspaces=false,
  columns=flexible,
  basicstyle={\ttfamily\scriptsize},
  numbers=none,
  numberstyle=\tiny\color{gray},
  keywordstyle=\color{blue},
  commentstyle=\color{dkgreen},
  stringstyle=\color{mauve},
  breaklines=true,
  numbers=left,
  numbersep=5pt,
  upquote=true,
  breakatwhitespace=true,
  tabsize=2,
  captionpos=b,
  keywordstyle=[2]{\color{cerulean}},
  morekeywords=[2]{assume, assert, int_nd, bool_nd, is_deref, valid_access, alloc, gep, store, load, havoc, read, write, malloc, input, sanitize, output, scanf, printf},
  keywordstyle=[3]{\color{orange}},
  morekeywords=[3]{@len, @char, @cap, @buf},
  keywordstyle=[4]{\color{blue}},
  morekeywords=[4]{NULL}
}

\lstdefinelanguage{yaml}{
  keywords={true,false,null},
  keywordstyle=\color{blue}\bfseries,
  sensitive=false,
  comment=[l]{\#},
  commentstyle=\color{green!50!black}\ttfamily,
  stringstyle=\color{red}\ttfamily,
  morestring=[b]',
  basicstyle=\ttfamily\scriptsize,
  frame=none,
  breaklines=true,
  columns=fullflexible,
}

\ProvideTextCommand{\textquotedbl}{OT1}{\char34}%

\newcommand{\stacklabel}[1]{\stackrel{\smash{\scriptscriptstyle \mathrm{#1}}}}
\newcommand{\Def}{\stacklabel{\mathrm{def}}}
\newcommand{\newwork}[1]{\textcolor{magenta}{#1}}

\newcommand{\prog}{\m{P}}

\newcommand{\statement}{\m{S}}
\newcommand{\condition}{\m{C}}
\newcommand{\arithmetic}{\m{A}}

\newcommand{\constant}{\m{Const}}
\newcommand{\op}{\m{op}}
\newcommand{\opint}{\op_{\mathit{int}}}

\newcommand{\opcmp}{\op_{\mathit{cmp}}}
\newcommand{\oplogic}{\op_{\mathit{logic}}}

\newcommand{\programp}{\mathit{pg}}
\newcommand{\arithe}{\mathit{a}}
\newcommand{\statements}{\mathit{s}}
\newcommand{\conditioncd}{\mathit{b}}

\newcommand{\pointerp}{\mathit{p}}
\newcommand{\pointerq}{\mathit{q}}
\newcommand{\varx}{\mathit{x}}
\newcommand{\inti}{\mathit{i}}
\newcommand{\constc}{\mathit{c}}
\newcommand{\elementsize}{\mathit{s}}

\newcommand{\kif}{\m{if}}
\newcommand{\kthen}{\m{then}}
\newcommand{\kelse}{\m{else}}
\newcommand{\kwhile}{\m{while}}

\newcommand{\true}{\m{true}}

\newcommand{\makeref}{\m{alloc}}

\newcommand{\kinput}{\m{input}}
\newcommand{\koutput}{\m{output}}
\newcommand{\ksanitize}{\m{sanitize}}

\newcommand{\tool}[1]{\mbox{\textsc{#1}}\xspace}
\newcommand{\crab}{\tool{Crab}}
\newcommand{\seadsa}{\tool{SeaDSA}}
\newcommand{\teadsa}{\tool{TeaDSA}}
\newcommand{\seahorn}{\tool{SeaHorn}}

\newcommand{\comptaint}{\tool{CompTaint}}
\newcommand{\ptaint}{\tool{P/Taint}}
\newcommand{\phasar}{\tool{PhASAR}}
\newcommand{\svf}{\tool{SVF}}
\newcommand{\flowdroid}{\tool{FlowDroid}}
\newcommand{\clangtaint}{\tool{ClangTaint}}
\newcommand{\mopsa}{\tool{Mopsa-Nexp}}
\newcommand{\expone}{\tool{Baseline}}
\newcommand{\exptwo}{\tool{Pta}}
\newcommand{\expthree}{\tool{Taint}}

\newcommand{\cenv}{\rho}
\newcommand{\heap}{h}
\newcommand{\taintMap}{T}

\newcommand{\state}{\sigma}
\newcommand{\stateset}{\Sigma}

\newcommand{\srcset}{S}

\newcommand{\tgtset}{D}

\newcommand{\abssrcset}{\srcset^\sharp}
\newcommand{\abstgtset}{\tgtset^\sharp}

\newcommand{\lfp}{\m{lfp}}

\newcommand{\crabir}{CrabIR\xspace}
\newcommand{\code}[1]{\lstinline[basicstyle=\footnotesize\ttfamily] {#1}\xspace}

\crefname{cline}{line}{lines}
\crefname{irline}{line}{lines}

\newcommand{\nnum}{{\mathbb N}\xspace}
\newcommand{\nint}{{\mathbb Z}\xspace}
\newcommand{\valset}{{\mathcal X}\xspace}
\newcommand{\ptrvalset}{\valset_{\mathcal P}\xspace}
\newcommand{\val}{\mathit{val}\xspace}
\newcommand{\numval}{\mathit{n}\xspace}
\newcommand{\addrval}{\mathit{a}\xspace}
\newcommand{\basefld}{\mathit{b}\xspace}
\newcommand{\offsetfld}{\mathit{o}\xspace}
\newcommand{\varset}{{\mathcal V}\xspace}

\newcommand{\varptrset}{{\mathcal V}_{\mathcal P}\xspace}
\newcommand{\varintset}{{\mathcal V}_{\mathcal I}\xspace}

\newcommand{\nodeset}{N\xspace}
\newcommand{\node}{n\xspace}
\newcommand{\cellset}{C\xspace}
\newcommand{\cell}{c\xspace}
\newcommand{\of}{o\xspace}
\newcommand{\ofset}{\nnum}
\newcommand{\oftstart}{s\xspace}
\newcommand{\oftend}{e\xspace}
\newcommand{\edgeset}{E\xspace}
\newcommand{\edge}{e\xspace}
\newcommand{\graph}{g\xspace}
\newcommand{\graphset}{G\xspace}

\newcommand{\envmap}{E_{\mathcal P}\xspace}

\newcommand{\newdsa}{I-DSA\xspace}
\newcommand{\dsabin}{\tool{DSAbin}}

\newcommand{\clean}{\mathsf{clean}}
\newcommand{\tainted}{\mathsf{tainted}}
\newcommand{\taintLattice}{\mathcal{T}}
\newcommand{\taintleq}{\sqsubseteq_\taintLattice}
\newcommand{\taintjoin}{\sqcup_\taintLattice}
\newcommand{\taintstate}{\state_t}
\newcommand{\taintstateset}{\stateset_t}
\NewDocumentCommand{\bigtaintjoin}{o}{%
  \mathop{\bigsqcup}\limits_{\IfNoValueTF{#1}{\taintLattice}{#1}}%
}

\NewDocumentCommand{\taintlookup}{o}{\hat{\taintMap}\IfNoValueTF{#1}{ }{(#1)}}
\newcommand{\prop}{\xrightarrow{\taintMap}}
\newcommand{\propshort}[2]{#1\prop#2}

\newcommand{\absmemstate}{\state^\sharp}
\newcommand{\abstaintstate}{\taintMap^\sharp}
\newcommand{\absprop}{\xrightarrow{\abstaintstate}}
\newcommand{\abspropshort}[2]{#1\absprop#2}

\newcommand{\absstate}{\taintstate^\sharp}

\newcommand{\locs}{{\mathcal L}\xspace}
\newcommand{\loc}{l\xspace}
\newcommand{\abslocs}{\locs^\sharp}
\newcommand{\absloc}{\loc^\sharp}

\newcommand{\absofset}{O^\sharp}
\newcommand{\absof}{\of^\sharp}

\newcommand{\tran}[2]{\llbracket#1\rrbracket_{\scriptscriptstyle #2}}

\renewcommand{\gets}{\mathbin{=}}

\algdef{SE}[FUNCTION]{MyFunction}{EndMyFunction}[2]{#1(#2)}{\textbf{end}}
\algtext*{EndMyFunction}%
\algnewcommand\FunctionNoNumber[2]{\Statex\MyFunction{#1}{#2}}

\algnewcommand\algorithmicelsif{\textbf{elif}}

\algdef{C}[IF]{IF}{ElsIf}[1]{\algorithmicelsif\ #1}
\algrenewcommand\algorithmicforall{\textbf{foreach}}

\algdef{SE}[FOR]{ForAll}{EndForAll}[1]{\algorithmicforall\ #1\ }{\algorithmicend\ \algorithmicforall}
\algtext*{EndForAll}
\algdef{SE}[FUNCTION]{NewFunction}{EndNewFunction}[2]{\textcolor{magenta}{#1(#2)}}{\textbf{end}}
\algtext*{EndNewFunction}%
\algnewcommand\New[1]{\State \textcolor{magenta}{#1}}
\algnewcommand\Newx[1]{\Statex \textcolor{magenta}{#1}}

\newcommand{\concretetran}{\tran{\statements}{}}
\newcommand{\concretetranon}[1]{\tran{#1}{}}

\newcommand{\dsatran}{\tran{\statements}{D}^\sharp}
\newcommand{\dsatranon}[1]{\tran{#1}{D}^\sharp}
\newcommand{\taintctran}{\tran{\statements}{T}}

\newcommand{\dfatran}{\tran{\statements}{T}^\sharp}
\newcommand{\dfatranon}[1]{\tran{#1}{T}^\sharp}

\newcommand{\memtran}{\tran{\statements}{m}^\sharp}

\newcommand{\unifyCells}{\mt{unifyCells}}
\newcommand{\unifyCellsSame}{\mt{unifyCellsSame}}
\newcommand{\unifyCellsOther}{\mt{unifyCellsOther}}

\newcommand{\mergeCells}{\mt{mergeCells}}
\newcommand{\unifyNodes}{\mt{unifyNodes}}

\newcommand{\isCollapsed}{\mt{isCollapsed}}

\newcommand{\lookUp}{\mt{find}}

\newcommand{\redirectEdges}{\mt{redirectEdges}}

\newcommand{\mergeOrNew}{\mt{mergeOrNew}}

\newcommand{\eval}{\mt{eval}}

\newcommand{\kills}{\mt{kill}}
\newcommand{\abskills}{\kills^\sharp}

\newcommand{\m}[1]{\mathsf{#1}}

\newcommand{\mt}[1]{\mathtt{#1}}

\newcommand{\dom}{\mt{dom}}
\newcommand{\powerset}{\mathcal{P}}

\newcommand{\update}[2]{#1[#2]}
\newcommand{\fresh}{\mathit{f}}
\newcommand{\freshnode}{\node_{\fresh}}

\newcommand{\getvars}{\mt{vars}}
\newcommand{\overlap}{\mt{overlap}}
\newcommand{\concretdom}{{\mathbb D}}
\newcommand{\taintdom}{\concretdom_t}
\newcommand{\absdomain}{\concretdom^\sharp}
\newcommand{\abstaintdom}{\absdomain_t}

\newcommand{\absmemdom}{\absdomain_m}
\newcommand{\abssetdom}{\absdomain_s}
\newcommand{\numdom}{\absdomain_n}

\newcommand{\membot}{\bot_m}
\newcommand{\memjoin}{\sqcup_m}
\newcommand{\memleq}{\sqsubseteq_m}

\newcommand{\concretization}{\gamma}
\newcommand{\concretegraph}{\concretization_{dsa}}

\newcommand{\concretemem}{\concretization_{m}}

\newcommand{\concretetaint}{\concretization_{t}}
\newcommand{\taintir}{\mbox{C-IR}\xspace}

\algrenewcommand\algorithmicindent{1em} 
\newif\ifanonymous
\anonymousfalse%
\newcommand{\appref}[1]{Appendix~\ref{#1}}
\begin{document}
\title{Taint Analysis via Heap-Aware Propagation}
\ifanonymous
\author{Anonymous}
\authorrunning{Anonymous}
\else
\author{Yusen Su\inst{1,2}\orcidID{0009-0004-8813-0797} \and
Jorge A. Navas\inst{3}\orcidID{0000-0002-0516-1167} \and
\mbox{Arie Gurfinkel}\inst{1}\orcidID{0000-0002-5964-6792}}
\institute{Department of Electrical and Computer Engineering, University of Waterloo, Waterloo, ON, Canada \and
Black Duck Software, Inc., Burlington, MA, USA \and
Certora Inc., Seattle, WA, USA}
\authorrunning{Y. Su et al.}
\fi
\maketitle%
\setcounter{footnote}{0}
\begin{abstract}

Taint analysis for C depends on how precisely the analysis tracks tainted values as they flow through memory objects. Without field sensitivity, a tainted object field can spuriously taint an entire object, producing significant false alarms. In this paper, we present a field-sensitive taint analysis that combines Data Structure Analysis (DSA) for field-sensitive points-to information with a data-flow analysis for tracking explicit dependencies. We identify that existing DSA-style analyses lose precision on a common object pattern: a C struct object that mixes scalar fields with a flexible array member. When such an array is accessed with symbolic indices, DSA loses points-to field sensitivity for the entire object, increasing false alarms in the data-flow analysis.

To address this, we propose \newdsa{}, which extends DSA by representing field positions as intervals, so that a pointer with symbolic bounds merges only the fields within those bounds. We further develop a sound data-flow analysis based on abstract interpretation. We implement \newdsa{} on top of \seadsa{} and the data-flow analysis inside \crab{}. Our tool targets LLVM IR and supports configurable taint policies by instrumenting the IR with taint intrinsics. Our experiments show that \newdsa{} reduces the false alarms caused by this pattern at little cost in the pointer analysis, that \svf{}, an inclusion-based pointer analysis, shows a similar field-sensitivity loss, and that our taint analysis scales better than \phasar{} on real-world programs. 
\end{abstract}
\section{Introduction}
\label{sec:intro}
Modern software systems, including network utilities, databases, and serialization libraries, continuously process external inputs and API requests. These interactions increase the risk of unintended information leaks~\cite{cisco_cve_2026_20128} or malicious data breaches~\cite{nist_cve_2025_14847,nist_cve_2025_40551}. Taint analysis~\cite{DBLP:conf/pldi/ArztRFBBKTOM14,DBLP:conf/issta/HuangDMD15,DBLP:journals/corr/abs-2601-17390} addresses this risk by checking whether untrusted or sensitive data can reach critical sinks, such as diagnostic logging, where private information may be exposed.

A main challenge of taint analysis is the high rate of false alarms caused by over-approximation of taint flows, which leads to spurious warnings and unreliable leak detection. One source of this imprecision is the lack of \emph{field sensitivity}. In C, data is often organized into complex memory objects such as structs and arrays, and taint often affects a specific field rather than the entire object. Without field sensitivity, taint analysis may treat a whole object as tainted because it cannot distinguish between tainted and untainted fields within the same object. Existing work such as \comptaint~\cite{DBLP:conf/sigsoft/BanerjeeCEFHLLP23} and \ptaint~\cite{DBLP:journals/pacmpl/GrechS17} shows that precision depends heavily on the underlying memory abstraction. Integrating a pointer analysis that preserves field sensitivity therefore improves precision.

In this paper, our taint analysis combines Data Structure Analysis~\cite{DBLP:conf/sas/GurfinkelN17,DBLP:conf/pldi/LattnerA05,DBLP:conf/pldi/LattnerLA07} (DSA) with data-flow analysis (DFA): DSA first infers points-to information, and DFA then uses it for memory-aware taint propagation. To improve precision, we introduce a new DSA variant with interval-based offsets, allowing taint analysis to preserve more field-level precision than when using standard DSA. We build the DFA by abstract interpretation~\cite{DBLP:conf/popl/CousotC77} to soundly over-approximate the explicit taint flows of the program. Together, these analyses make taint propagation more precise while preserving practical scalability.

We implement our taint analysis on top of the \seahorn~\cite{DBLP:conf/tacas/GurfinkelKN15} verification framework, targeting LLVM IR~\cite{DBLP:conf/cgo/LattnerA04}. Our new DSA extends the implementation of \seadsa{}~\cite{DBLP:conf/sas/GurfinkelN17}, and our DFA is implemented in \crab{}~\cite{DBLP:conf/vstte/GurfinkelN21}. Our tool takes taint policies in YAML and instruments LLVM IR with taint intrinsics. We evaluate our approach in three experiments on $197$ benchmarks, including $154$ real-world programs. First, we run the same taint analysis with \seadsa{} and our new DSA. Our DSA reports $3.9\%$ fewer warnings, entirely due to recovered field sensitivity, while the total running time increases by $0.8\%$. Second, we compare our DSA with \svf{}'s inclusion-based pointer analyses on the same alias queries. \svf{} also loses field sensitivity on a pattern similar to the one in \cref{sec:overview}. We suspect that other pointer analyses share this limitation. Third, we compare our taint analysis with \phasar{} under a shared policy. Their results differ mainly because of different propagation rules and memory models. Our DSA and taint analysis scale better on real-world programs than \svf{} and \phasar{}, respectively.

We make three contributions: we formalize the semantics of explicit-flow taint analysis; we design a pointer-analysis abstraction that recovers field sensitivity that existing DSA-style approaches lose; and we implement our taint analysis.

\cref{sec:overview} presents an overview, \cref{sec:concrete_lan_sem} the concrete semantics, \cref{sec:dsa_o} our new DSA, \cref{sec:taint_sem} the data-flow analysis, \cref{sec:impl} the implementation, \cref{sec:eval} the evaluation, and \cref{sec:related} related work. \section{Overview}
\label{sec:overview}
This section motivates our approach with a simplified example from a real-world codebase and shows where previous approaches lose precision.

\newcommand{\rb}{\mathit{rb}}
\begin{figure}[t]
\lstset{escapeinside={(*@}{@*)}}
\begin{minipage}[b]{0.52\linewidth}
\vspace{0pt}
\begin{lstlisting}[basicstyle=\footnotesize, language=C, style=mystyle, aboveskip=0pt, belowskip=0pt, xleftmargin=11.81pt]
typedef struct repl_block {
  int id;
  size_t size;
  size_t used;
  int buf[]; // flexible array member
} repl_block_t;
static int block_id = 0;
void fRepBuf(const int *s, size_t len) {  (*@\label[cline]{cline:rdsrcs}@*)
  repl_block_t *b;
  size_t copy;
  b = malloc(sizeof(repl_block_t) + len * 4);(*@\label[cline]{cline:rdballoc}@*)
  b->id = block_id++; (*@\label[cline]{cline:rdidassign}@*)
  b->size = len; (*@\label[cline]{cline:rdszassign}@*)
  b->used = 0; (*@\label[cline]{cline:rdusedinit}@*)
  size_t rem = b->size - b->used;
  copy = len < rem ? len : rem;
  for (size_t i = 0; i < copy; ++i) {
    b->buf[i] = s[i]; (*@\label[cline]{cline:rdbufassign}@*)
  }
  b->used += copy; (*@\label[cline]{cline:rdusedassign}@*)
  log("block id=%d used=%zu size=%zu",
      b->id, b->used, b->size); (*@\label[cline]{cline:rdsink}@*)
}
\end{lstlisting}
\subcaption{}
\label{fig:redis_code}
\end{minipage}
\hfill
\begin{minipage}[b]{0.47\linewidth}
\vspace{0pt}
\centering
\scalebox{.9}{%
\begin{tikzpicture}[
    y=8.856pt, >=Latex,
    line width=0.45pt,
    line cap=round,
    line join=round,
    proc/.style={
        draw, fill=gray!55,
        text=white,
        rounded corners=2pt,
        inner xsep=4pt, inner ysep=1.5pt,
        minimum width=15mm,
        minimum height=3mm,
        font=\ttfamily\small,
        align=center
    },
    oval/.style={
        draw,
        ellipse,
        minimum width=8mm,
        minimum height=8mm,
        inner xsep=0pt, inner ysep=0pt,
        font=\ttfamily\scriptsize,
        align=center
    },
    lab/.style={
        font=\sffamily\scriptsize
    },
    flow/.style={->}
]
\node[proc]  (srcs)   at (0,-8) {source};%
\node[lab, above=1mm of srcs] {\texttt{s}};
\node[proc]  (srclen) at (2.0,-8) {source};
\node[lab, above=1mm of srclen] {\texttt{len}};
\node[proc] (setsize)  at (2.0,-13) {field store};%
\node[proc] (storebuf) at (0,-18) {array store};%
\node[proc] (updused)  at (2.0,-20) {field store};%
\node[proc] (sink2)    at (2.0,-22.2) {sink};%
\def\bh{2.0}
\coordinate (memsw) at (0,-5.8);
\draw ($(memsw)+(0,\bh)$) -- ++(5,0.0);
\draw ($(memsw)$) -- ++(5,0.0);
\draw ($(memsw)$) -- ++(0,\bh);
\draw[dashed] ($(memsw)+(5,0.0)$) -- ++(0,\bh);
\draw ($(memsw)+(1.1,0)$) -- ++(0,\bh);
\draw ($(memsw)+(2.2,0)$) -- ++(0,\bh);
\draw ($(memsw)+(3.3,0)$) -- ++(0,\bh);
\coordinate (fid)   at ($(memsw)+(0,\bh)$);
\coordinate (fsize) at ($(memsw)+(1.1,\bh)$);
\coordinate (fused) at ($(memsw)+(2.2,\bh)$);
\coordinate (fbuf)  at ($(memsw)+(3.3,\bh)$);
\draw[decorate, decoration={brace, amplitude=4pt}] (fbuf) -- ($(memsw)+(5,\bh)$);
\node[lab] (of0) at ($(fid)+(-0.3,0.0)$) {0};
\node[lab] (of4) at ($(fsize)+(-0.16,0.45)$) {4};
\node[lab] (of12) at ($(fused)+(-0.2,0.45)$) {8};
\node[lab] (of20) at ($(fbuf)+(-0.2,0.45)$) {12};
\node[lab] (noden) at ($(fid)+(-0.7,-1.0)$) {\texttt{node }$\rb$};
\node[lab] (oid)   at ($(fid)+(0.5,-0.34)$)   {\texttt{id}};
\node[lab] (osize) at ($(fsize)+(0.6,-0.34)$) {\texttt{size}};
\node[lab] (oused) at ($(fused)+(0.6,-0.34)$) {\texttt{used}};
\node[lab] (obuf)  at ($(fbuf)+(0.9,-0.34)$)  {\texttt{buf[]}};
\node[oval] (b)     at ($(fid)+(-0.5,2.0)$)   {\texttt{b}};
\node[oval] (bid)   at ($(fid)+(0.5,2.0)$)    {\texttt{bid}};
\node[oval] (bsize) at ($(fsize)+(0.6,2.0)$)  {\texttt{bsize}};
\node[oval] (bused) at ($(fused)+(0.6,2.0)$)  {\texttt{bused}};
\node[oval] (bbuf)  at ($(fbuf)+(0.9,2.0)$)   {\texttt{bbufi}};
\draw[flow] (bid.south)   -- (fid);
\draw[flow] (b.south)   -- (fid);
\draw[flow] (bsize.south) -- (fsize);
\draw[flow] (bused.south) -- (fused);
\draw[-] (bbuf.south)  -- ($(memsw)+(4.15,\bh+0.27)$);
\draw[flow] (srcs.south) -- (storebuf.north);
\draw[flow] (storebuf.east) .. controls +(3.9,0) and +(0,-6) .. ($(fbuf)+(0.7,-\bh)$);
\draw[flow] (srclen.south) -- (setsize.north);
\draw[flow] (setsize.west) .. controls (0.8,-12.0) and (1.0,-7.3) .. (1.35,-5.8);
\draw[flow] (setsize.south) -- (updused.north);
\draw[flow,dashed] (updused.south) -- (sink2.north);
\draw[flow] (updused.east) .. controls +(1.1,0) and +(0.4,-7) .. ($(fused)+(0.6,-\bh)$);
\path (0,-23.6);
\end{tikzpicture}%
}
\subcaption{}
\label{fig:flow_redis}
\end{minipage}
\caption{(a) A program, and (b) its taint-flow graph over the data structure graph.}
\label{fig:empip}
\end{figure} 
Consider the example~\cite{redis-replication} in \cref{fig:redis_code}. A \code{struct repl_block} object referenced by \code{b} is allocated at \cref{cline:rdballoc}. Its allocation size depends on the user-provided parameter \code{len} (\cref{cline:rdsrcs}) because \code{struct repl_block} ends with a flexible array member \code{buf[]}. This array occupies extra bytes allocated beyond the static structure size. The code then initializes \code{b->buf} from user-controlled inputs under the bound \code{i < copy}. Finally, the program calls \code{log} to report block information.

Clearly, the field \code{id} is not influenced by user input. At \cref{cline:rdidassign}, its value is derived from the global serial number \code{block_id}. Showing this fact requires two ingredients: (1) alias information proving that no other pointer may access \code{b->id}, and (2) data-flow information showing that user-controlled inputs are not propagated into this field. We therefore combine a points-to analysis (PTA) to identify possible memory targets with a taint data-flow analysis (DFA) to propagate taint across assignments and memory operations.

PTA computes alias information from the memory operations of a program, indicating which objects each pointer may reference and how pointers are related. Our PTA builds on Data Structure Analysis~\cite{DBLP:conf/pldi/LattnerA05,DBLP:conf/pldi/LattnerLA07} (DSA), which represents aliasing relationships with a data structure graph (DSG). Like a points-to graph, a DSG is a directed graph whose edges represent points-to relations. One difference is that a DSG also models the fields within each object. Each node is an abstract memory object (e.g., a group of objects allocated at the same allocation site), and its fields are represented as separate cells at distinct byte offsets. Since the graph records which pointer accesses which object field, DSA is field-sensitive. In our example, the DSG that our PTA computes (top of \cref{fig:flow_redis}) contains a node $\rb$ for the object pointed to by \code{b}, with a distinct field \code{id} at offset \(0\). Other fields are represented at different offsets.

A DFA for taint computes how values from external sources (e.g., user or file inputs) propagate to sensitive operations that may expose data (e.g., logging or system calls). Such untrusted inputs are modeled as \emph{sources}, and potential leakage points are modeled as \emph{sinks}. The analysis tracks taint propagation under explicit policies for assignments, memory loads/stores, and function calls. In our example, function input parameters are sources and logging calls are sinks. The taint-flow graph is shown in \cref{fig:flow_redis}. Assignments such as those at \cref{cline:rdidassign,cline:rdszassign} propagate taint along data dependencies from read locations to written locations. Concretely, \cref{cline:rdidassign} sets \code{id} from \code{block_id}. Together with DSA alias information, this implies that no user-controlled input reaches \code{id}. By contrast, fields such as \code{size}, \code{used}, and elements of \code{buf} may depend on tainted inputs. The sink at \cref{cline:rdsink} may therefore leak user-controlled inputs from \cref{cline:rdsrcs}.

However, the precision of a taint analysis depends on the field sensitivity of the underlying PTA, and DSA can lose field sensitivity on objects like the one in our example. The cause is the memory access at \cref{cline:rdbufassign}: the index \code{i} is symbolic, so \code{b->buf[i]} may access any element of the array. Existing DSA variants~\cite{DBLP:conf/sas/GurfinkelN17,DBLP:conf/pldi/LattnerA05,DBLP:conf/pldi/LattnerLA07} identify each field by a single constant offset, and \code{b->buf[i]} has no such offset. They therefore merge all offsets of $\rb$ into a single abstract location. Their graph has one location for \code{id}, \code{size}, \code{used}, and \code{buf}, whereas the graph in \cref{fig:flow_redis} has four fields. The store of \code{len} at \cref{cline:rdszassign} then also taints \code{id}, which produces a false alarm for \code{b->id} at the sink (\cref{cline:rdsink}). In this example, however, no access with a symbolic index targets \code{id}, so \code{id} should keep a separate field. We propose a refinement of DSA that represents offsets as intervals, so an access with a symbolic index merges only the fields whose ranges overlap and keeps the others separate (\cref{sec:dsa_o}).

Building on this refined memory abstraction, the DFA in \cref{sec:taint_sem} marks variables and memory locations as tainted or clean and propagates taint along explicit data dependencies induced by assignments, memory loads/stores, and function calls. To prune flows along paths that are provably infeasible, we also combine the DFA with a value analysis that tracks numerical invariants.

 \section{Language and Semantics}
\label{sec:concrete_lan_sem}
This section describes the target language for analysis and its concrete semantics.

\begin{figure}[t]
\centering
\resizebox{\textwidth}{!}{%
\begin{minipage}{1.145\textwidth}
\begin{alignat*}{3}
\prog \ni \programp \;::=&\; \statement^{*} &&\qquad &&\qquad \text{(Program)}\\
\statement \ni \statements  \;::=&\; \pointerp = \makeref(\inti) \mid \pointerp = \pointerq  + \of \mid \pointerp = \pointerq + (\inti \times \elementsize) \mid \varx = *\pointerp \mid *\pointerp = \varx &&\qquad &&\qquad \text{(Memory)}\\
&\mid \inti = \arithe \mid \varx =  \kinput() \mid \ksanitize(\varx) \mid \koutput(\varx) \mid \statements;\ \statements &&\qquad &&\qquad \text{(Arith., Taint, Seq.)}\\
&\mid \kif(\conditioncd)\ \kthen\ \statements\ \kelse\ \statements \mid \kwhile(\conditioncd)\ \statements &&\qquad &&\qquad \text{(Control Flow)}\\
\condition \ni \conditioncd \;::=&\; \arithe \; \opcmp \; \arithe \mid \neg~ \conditioncd \mid \conditioncd \; \oplogic \; \conditioncd \mid \pointerp == \pointerq \mid \pointerp \neq \pointerq &&\qquad &&\qquad \text{(Guards)}\\
\arithmetic \ni \arithe \;::=&\; \constc \in \constant \mid \inti \in \varintset \mid \arithe \; \opint \; \arithe &&\qquad &&\qquad \text{(Arith.\ Expressions)}
\end{alignat*}
\end{minipage}%
}
\caption{The syntax of \taintir.}
\label{fig:lang-syntax}
\vspace{-1em}
\end{figure} \newcommand{\ptr}[2]{(#1, #2)}
\newcommand{\addr}{\ell}%
\newcommand{\baddr}{b}

\newcommand{\pol}{\mathcal{P}}
\newcommand{\polalloc}{\pol_{\text{alloc}}}
\newcommand{\guard}[1]{\mathsf{grd}_{#1}}

\begin{figure}[t]
\centering
\small
\[
\begin{array}{@{}r@{\;}l@{\;\;}r@{}}
\concretetranon{\pointerp = \makeref(\inti)}(\cenv, \heap, \newwork{\taintMap}) =& (\cenv[\pointerp \mapsto \ptr{\baddr}{0}],\, \heap',\, \newwork{\propshort{\{\}}{\pointerp}}) & (\baddr, \heap') = \heap.\makeref(\cenv(\inti)) \\
\concretetranon{\inti = \arithe}(\cenv, \heap, \newwork{\taintMap}) =& (\cenv[\inti \mapsto \cenv.\eval(\arithe)],\, \heap,\, \newwork{\propshort{\getvars(\arithe)}{\inti}}) &  \\
\concretetranon{\pointerp = \pointerq + \of}(\cenv, \heap, \newwork{\taintMap}) =& (\cenv[\pointerp \mapsto \ptr{\baddr}{\of_\pointerq + \of}],\, \heap,\, \newwork{\propshort{\pointerq}{\pointerp}}) & \ptr{\baddr}{\of_\pointerq} = \cenv(\pointerq) \\
\concretetranon{\pointerp = \pointerq + (\inti \times \elementsize)}(\cenv, \heap, \newwork{\taintMap}) =& (\cenv[\pointerp \mapsto \ptr{\baddr}{\of_\pointerq + \cenv(\inti) * \elementsize}],\, \heap,\, \newwork{\propshort{\pointerq}{\pointerp}}) & \ptr{\baddr}{\of_\pointerq} = \cenv(\pointerq) \\
\concretetranon{\varx = \kinput()}(\cenv, \heap, \newwork{\taintMap}) =& (\cenv[\varx \mapsto \val],\, \heap,\, \newwork{\taintMap \cup \{\varx\}}) & \val = \kinput() \\
\concretetranon{\ksanitize(\varx)}(\cenv, \heap, \newwork{\taintMap}) =& (\cenv,\, \heap,\, \newwork{\taintMap \setminus \{\varx\}}) & \\
\concretetranon{\varx = *\pointerp}(\cenv, \heap, \newwork{\taintMap}) =& (\cenv[\varx \mapsto \heap(\addrval_\pointerp)],\, \heap,\, \newwork{\propshort{\addrval_\pointerp}{\varx}}) & \addrval_\pointerp = \cenv(\pointerp) \\
\concretetranon{*\pointerp = \varx}(\cenv, \heap, \newwork{\taintMap}) =& (\cenv,\, \heap[\addrval_\pointerp \mapsto \cenv(\varx)],\, \newwork{\propshort{\varx}{\addrval_\pointerp}}) & \addrval_\pointerp = \cenv(\pointerp)
\end{array}
\]
\[
\begin{array}{@{}l@{}}
\concretetranon{\statements}(X) = \{\concretetran(\state) \mid \state \in X\} \text{ for primitive } \statements, \qquad \concretetranon{\statements_1;\ \statements_2}(X) = \concretetranon{\statements_2}(\concretetranon{\statements_1}(X)),\\
\concretetranon{\kif(\conditioncd)\ \kthen\ \statements_t\ \kelse\ \statements_f}(X) = \concretetranon{\statements_t}(\guard{\conditioncd}(X)) \cup \concretetranon{\statements_f}(\guard{\neg\conditioncd}(X)),\\
\concretetranon{\kwhile(\conditioncd)\ \statements}(X) = \guard{\neg\conditioncd}\bigl(\lfp\, \lambda Y.\;X \cup \concretetranon{\statements}(\guard{\conditioncd}(Y))\bigr)
\end{array}
\]
\caption{Concrete collecting semantics. \newwork{\cref{sec:taint_sem} explains the highlighted parts}.}
\label{fig:trans}
\end{figure} 
We lower C programs to a simple intermediate representation (IR), called \taintir, shown in \cref{fig:lang-syntax}. Compound C statements are translated into sequences of primitive statements in \taintir. We define a \taintir{} statement \emph{primitive} if it is not a sequence, a branch, or a loop.
We assume the input language is a subset of C with integer variables \( \inti \in \varintset \) and pointer variables \( \pointerp, \pointerq \in \varptrset \), where \( \varx \in \varset = \varintset \cup \varptrset \). Pointers range only over heap-object addresses (no stack-address values).
\taintir supports dynamic allocation, integer arithmetic, and pointer arithmetic with either fixed non-negative byte offsets ($\of\in\nnum$) or indexed byte offsets ($\inti \times \elementsize$) under a constant stride $\elementsize\in\nnum^*$. Load/store statements model memory access, while taint primitives (\code{input}, \code{sanitize}, and \code{output}) identify taint sources, sanitizers, and sinks. All three statements act on variables, and memory contents are tainted or cleaned only through stores.

\taintir{} is a presentation IR used to define semantics concisely, while the implementation analyzes LLVM IR. In \taintir{}, a value is either an integer or a pointer, and there is no cast between the two. For simplicity, \taintir{} also restricts pointer arithmetic to non-negative offsets and omits function calls and deallocation. The implementation has no such restrictions.

We model memory as a collection of objects, where each allocation creates a fresh, unique object.
We assume that memory is word-addressable~\cite{DBLP:conf/vstte/GurfinkelN21}: an object consists of non-overlapping fields, every field has the size of one word and stores one value, and every load and store accesses exactly one field. A pointer is a pair of a base address and a byte offset. The word size is a fixed parameter, and all offsets are multiples of it, so loads and stores have no access size. For presentation, we fix the word size to $4$ bytes, as on a 32-bit target where \code{int}, \code{size_t}, and pointers take $4$ bytes. For example, \code{\&b->size} in \cref{fig:redis_code} is written in \taintir{} as \code{bsize = b + 4}, and \code{\&b->buf[i]} as \code{bbuf = b + 12;} \mbox{\code{bbufi = bbuf + (i * 4)}}. The implementation allows fields of different sizes.

Formally, values are pairs in $\valset \Def= \nnum \times \nint$, representing either integers or addresses. By convention, an integer $\numval$ is encoded as $(0,\numval)$, and an address $\addrval$ is encoded as $(\basefld,\offsetfld)$ with nonzero base $\basefld$.
For example, $3$ is represented as $(0,3)$, while $(10_{16},4)$ denotes offset $4$ of the object at base address $10_{16}$.
For brevity, we write $\addrval.\basefld$ and $\addrval.\offsetfld$ for the base and offset of an address, use $\ptrvalset$ for the set of address values, and define $\locs \Def= \varset \cup \ptrvalset$ as the set of memory locations.
A memory state $\state \in \stateset$ is a pair of an \emph{environment} $\cenv$ that maps variables to values, $\varset \to \valset$, and a \emph{heap} $\heap$ that maps addresses to values, $\ptrvalset \to \valset$.

\newcommand{\concretestep}{\mt{step}}

We define the concrete semantics of \taintir{} by a state transformer $\concretetran: \stateset \to \stateset$ for primitive statements (\cref{fig:trans}). The initial state consists of empty maps. To simplify memory modeling, we assume that $\heap$ provides an allocation operation $\makeref$ that always succeeds and returns a fresh base address. We also assume that $\cenv$ provides an evaluation function $\eval$ for arithmetic expressions, and write $\getvars(\arithe)$ for the variables in $\arithe$. Invalid dereference and invalid load/store behavior are outside the scope of this semantics. The collecting semantics lifts the single-state transformer to a set transformer $\concretetran: \concretdom \to \concretdom$ over sets of states, $\concretdom \Def= \powerset(\stateset)$ (\cref{fig:trans}). For guards, we define $\guard{\conditioncd}(X) \Def= \{\state \in X \mid \concretetranon{\conditioncd}(\state) = \true\}$. For a while loop, we take the least fixpoint of the corresponding set transformer.

 \section{Data Structure Analysis with Interval Offsets}
\label{sec:dsa_o}
\newcommand{\offsetdom}{\mathcal{O}}
\newcommand{\offsetleq}{\sqsubseteq_{\offsetdom}}
\newcommand{\offsetjoin}{\sqcup_{\offsetdom}}
\newcommand{\offsetmeet}{\sqcap_{\offsetdom}}
\newcommand{\offsetbot}{\bot_{\offsetdom}}
\newcommand{\offsettop}{\top_{\offsetdom}}
\newcommand{\offsetgamma}{\gamma_{\offsetdom}}
\newcommand{\offsetabstract}{\alpha_{\offsetdom}}
\newcommand{\offsetadd}{+_{\offsetdom}}
\newcommand{\offsetsub}{-_{\offsetdom}}
\newcommand{\offsetfunc}[1]{#1.\absof\xspace}
\newcommand{\pointsto}{\rightharpoonup}
\newcommand{\omin}[1]{#1.\oftstart\xspace}
\newcommand{\omax}[1]{#1.\oftend\xspace}

\newcommand{\nodefunc}[1]{#1.\node\xspace}

This section introduces \newdsa, a new DSA that abstracts cell offsets by intervals, whereas the original DSA uses constants. We first define our graph representation, then identify where the original DSA loses precision, and finally formalize the interval-based offset abstraction.

DSA~\cite{DBLP:conf/sas/GurfinkelN17,DBLP:conf/pldi/LattnerA05,DBLP:conf/pldi/LattnerLA07} summarizes memory objects and their points-to relations in a DSG (\cref{sec:overview}). A DSG contains abstract \emph{nodes}, each summarizing one or more memory objects, and each node contains abstract \emph{cells} for the distinct fields of these objects. A field with a statically known location is identified by its numeric offset. Edges between cells encode memory-to-memory points-to relations, and a separate map from pointer variables to cells records variable points-to facts.

Formally, a DSG is a tuple \(\graph = \langle \nodeset, \cellset, \edgeset, \envmap \rangle \in \graphset\), where \(\nodeset\) is the set of nodes, \(\absofset\) is the set of cell offsets, \(\cellset \subseteq \nodeset \times \absofset\) is the finite set of cells, a cell \(\cell = (\node, \absof)\) representing the fields that start at an offset in \(\absof\) within node \(\node\), \(\edgeset\) is the set of edges \(\edge\Def=\cell_1 \pointsto \cell_2\), mapping from a source cell to a target cell, and \(\envmap: \varptrset \pointsto \cellset\) is a partial mapping from pointer variables to cells. For later use, we write \(\edgeset[\cell_2 / \cell_1]_{\mathrm{key}}\) to replace the key \(\cell_1\) with \(\cell_2\), and \(\envmap[\cell_2 / \cell_1]_{\mathrm{value}}\) (or \(\edgeset[\cell_2 / \cell_1]_{\mathrm{value}}\)) to replace every occurrence of \(\cell_1\) with \(\cell_2\) in the map values, and \(\graph[\cell_2/\cell_1]_{\mathrm{value}} \Def= \langle \graph.\nodeset, \graph.\cellset, \graph.\edgeset[\cell_2/\cell_1]_{\mathrm{value}}, \graph.\envmap[\cell_2/\cell_1]_{\mathrm{value}} \rangle\) for a graph. If the key or value does not occur, the corresponding replacement is a no-op.

Let $\cellset_\node \Def= \{\cell \in \cellset \mid \nodefunc{\cell} = \node\}$ denote the cells of node $\node$. A DSG is \emph{well-formed} if it meets the following conditions:
\begin{description}
    \item[C.1 (Disjointness)] $\forall \node \in \nodeset, \cell_1, \cell_2 \in \cellset_\node.~ \cell_1 \neq \cell_2 \Leftrightarrow \neg\overlap(\offsetfunc{\cell_1}, \offsetfunc{\cell_2})$;%
    \item[C.2 (Collapsed)] $ \isCollapsed(\node) \Rightarrow |\cellset_\node| = 1 $;%
    \item[C.3 (Univalent)] $\forall \cell \in \dom(\edgeset).~ |\edgeset(\cell)| = 1$.
\end{description}
By C.1, the cells of a node are disjoint, so two pointers into it may alias only if they point to the same cell. By C.2, a \emph{collapsed} node has a single cell, so all pointers into it may alias. Condition C.3 and the partial map $\envmap$ enforce the invariant from unification-based PTA~\cite{DBLP:conf/popl/Steensgaard96} that each pointer/cell has at most one outgoing edge. As DSA represents the offset of each cell as a constant, $\overlap$ is defined by numeric comparison and $\isCollapsed$ is a special node property that is set to true when the field sensitivity is lost.

In our example in \cref{fig:redis_code}, the DSG at the top of \cref{fig:flow_redis} includes a node $\rb$ that separates the constant-offset fields referred to by pointers \code{b}, \code{bid}, \code{bsize}, and \code{bused}. The access at \cref{cline:rdbufassign}, however, introduces a symbolic offset through \code{i}. Existing DSA works represent this node $\rb$ by collapsing all cells within it. This is sound but at the expense of losing field sensitivity for the node $\rb$.%

\newdsa addresses this limitation by representing each cell offset as an interval. An array access with a symbolic index is modeled by an interval $[\oftstart,\oftend]$ that over-approximates the accessed offsets. For example, the access \code{b->buf[i]} at \cref{cline:rdbufassign} is represented by $[12,+\infty]$, since the buffer begins at offset $12$ and the array size is statically unknown. The interval abstraction preserves field sensitivity for the part of the object not affected by this access (in our example, the first 12 bytes holding \code{id}, \code{size}, and \code{used}). When the intervals of two accesses overlap, the affected cells are merged, whereas cells outside the overlap remain separate. 

Formally, an interval offset $\absof$ is written as $[\oftstart,\oftend]$, where $\omin{\absof}$ is the start offset and $\omax{\absof}$ is the end offset. Following~\cite{CousotIntervals}, cell offsets range over an interval domain \(\offsetdom\), a complete lattice \((\absofset, \offsetleq, \offsetjoin, \offsetmeet, \offsetbot, \offsettop)\) with partial order \(\offsetleq\), join \(\offsetjoin\), and meet \(\offsetmeet\). The lattice includes bottom \(\offsetbot\) and top \(\offsettop = [0,+\infty]\), since offsets are non-negative. We assume predefined abstraction and concretization functions: \(\offsetabstract: \powerset(\ofset) \mapsto \absofset\) and \(\offsetgamma: \absofset \mapsto \powerset(\ofset)\). We also use \(\offsetadd\) over intervals for arithmetic operations: \([a, b] \offsetadd [c, d] \Def= [a+c, b+d]\).

In \newdsa, to meet the well-formedness conditions, we let $\overlap(\absof_1, \absof_2) \Def= \absof_1 \offsetmeet \absof_2 \neq \offsetbot$. A node is collapsed when it has only one cell whose offset is maximum, denoted as $\isCollapsed(\node) \Def= |\cellset_\node|=1 \land \forall \cell \in \cellset_\node.~\offsetfunc{\cell}=\offsettop$.

The concretization $\concretegraph$ relates an abstract graph to concrete program states. Its formalization requires an auxiliary witness that shows how memory objects are represented by the abstract graph. We therefore introduce a simulation relation $\mu$ between a program state $\state$ and a DSG $\graph$: a partial function $\ptrvalset \pointsto \graph.\cellset$ that maps addresses of $\state$ to abstract cells of $\graph$.
\newcommand{\objo}{\mathit{obj}}

Consider a concrete memory object $\objo$ in $\state.\heap$. Let $A$ be the field locations of $\objo$ that $\mu$ maps, and let $\node \in \graph.\nodeset$ be a node representing $\objo$. The relation $\mu$ maps every location in $A$ to a cell of $\cellset_\node$. We place $\objo$ at some offset $x_{\objo} \geq 0$ of $\node$: the field at offset $\addrval.\offsetfld$ of $\objo$ lies at offset $x_{\objo} + \addrval.\offsetfld$ of $\node$, and $\mu$ maps it to a cell whose abstract offset covers that offset:
\[
  \forall \addrval \in A.\; x_{\objo} + \addrval.\offsetfld \in \offsetgamma(\offsetfunc{\mu(\addrval)})
\]
Intuitively, $x_{\objo}$ is an \emph{anchor} that aligns the object with the node. In the example below, it is $0$. It is positive when, e.g., $\pointerp$ may point to $\objo_1$ or to the field at offset $4$ of $\objo_2$: both targets share one cell (C.3), so $\objo_1$ starts at offset $4$.

Consider the code example in \cref{fig:redis_code} and its DSG (\cref{fig:flow_redis}). Suppose the object at \cref{cline:rdballoc} is allocated at base address $100$ (we write $100 + \offsetfld$ for the address $(100, \offsetfld)$) with $28$ bytes in total, so the array has four elements. The fields are at $100$ for \code{id}, $104$ for \code{size}, $108$ for \code{used}, and $112+4i$ for \code{buf[i]} where $0\le i \le 3$. In the DSG, node $\rb$ has cells $\cell_{\mathit{id}}$, $\cell_{\mathit{size}}$, $\cell_{\mathit{used}}$, and $\cell_{\mathit{buf[]}}$ with offsets $[0,0]$, $[4,4]$, $[8,8]$, and $[12,+\infty]$, respectively. The simulation relation $\mu$ places the object at offset $0$ of $\rb$ and maps these addresses to their cells: $(100, \cell_{\mathit{id}}), (104, \cell_{\mathit{size}}), (108, \cell_{\mathit{used}}) \in \mu$ and $(112+4i, \cell_{\mathit{buf[]}}) \in \mu$ for all $i \in [0, 3]$.

Using $\mu$, we define when a graph $\graph$ simulates a state $\state$, written $\state \models_\mu \graph$. All points-to relations in $\state$ must be represented by the corresponding edges in $\graph$:
$$
\state \models_\mu \graph \Leftrightarrow \begin{cases}  \forall \pointerp \in \varptrset \cap \dom(\cenv) .\; (\cenv(\pointerp), \envmap(\pointerp)) \in \mu \\
\forall (\addrval, \addrval') \in \heap.\; \exists \cell \in \cellset.\; (\addrval, \cell) \in \mu \land \left(\addrval' \in \ptrvalset \Rightarrow (\addrval',\edgeset(\cell)) \in \mu \right)
\end{cases}
$$
The first case requires each pointer variable to point to a cell that represents its target, and the second requires each memory-to-memory points-to relation to be represented by an edge between the corresponding cells. 

The concretization function $\concretegraph: \graphset \mapsto \powerset(\stateset)$ maps a DSG $\graph$ to the set of all concrete states that $\graph$ simulates through some $\mu$:
\[
\concretegraph(\graph) \Def= \{\state \in \stateset ~|~ \exists \mu.~ \state \models_\mu \graph \}
\]

As DSA incrementally constructs the graph, a pointer/cell may temporarily have multiple outgoing edges. For example, a pointer $\pointerp$ may point to both $(\node_1, \absof_1)$ and $(\node_2, \absof_2)$. This violates the invariant C.3, which requires every pointer/cell to have at most one outgoing edge. DSA resolves this by \emph{unification}, which merges the two target cells into a new cell $(\node_3, \absof_3)$ and redirects $\pointerp$ to it so that it has a single target. We capture this step with the procedure $\unifyCells(\cell_1,\cell_2)$, which restores C.3 by unifying $\cell_1$ and $\cell_2$ and, recursively, the targets of their outgoing edges. Our procedures update $\graph$ in place. Unification preserves the simulation: if a call of $\unifyCells$ turns $\graph$ into $\graph'$, then $\concretegraph(\graph) \subseteq \concretegraph(\graph')$ (Lemma~1 in \appref{proof:dsa-soundness}).

\begin{figure}[t]
\lstset{escapeinside={(*@}{@*)}}
\begin{minipage}[t]{\linewidth}
\vspace{0pt}
\begin{minipage}[t]{0.51\linewidth}
\vspace{0pt}
\begin{lstlisting}[basicstyle=\footnotesize, language=C, style=mystyle, aboveskip=0pt, belowskip=0pt, breaklines=false, numbersep=3pt, xleftmargin=9.81pt]
struct in_addr { uint32_t s_addr; };
struct in6_addr  { uint32_t s6_addr[4]; };
struct ifaddrs{struct sockaddr *ifa_addr;};
struct sockaddr_in {
  uint32_t sin_family;
  struct in_addr sin_addr;
  uint32_t sin_zero[2];
};
struct sockaddr_in6 {
  uint32_t sin6_family;
  uint32_t sin6_flowinfo;
  struct in6_addr sin6_addr;
  uint32_t sin6_scope_id;
};
\end{lstlisting}
\end{minipage}
\hfill
\begin{minipage}[t]{0.48\linewidth}
\vspace{0pt}
\begin{lstlisting}[basicstyle=\footnotesize, language=C, style=mystyle, firstnumber=15, aboveskip=0pt, belowskip=0pt, breaklines=false, numbersep=3pt, xleftmargin=2.5pt]
int if2ip(int af, char *interf, char *buf) {
  struct ifaddrs *iface;
  ...
  if (getifaddrs(&iface) >= 0) {
    void *addr;
    if (af == AF_INET6)
      addr = &((struct sockaddr_in6 *) (*@\label[cline]{cline:ipaddr1}@*)
                iface->ifa_addr)->sin6_addr;
    else
      addr = &((struct sockaddr_in *) (*@\label[cline]{cline:ipaddr2}@*)
                iface->ifa_addr)->sin_addr;
    ...
  }
}
\end{lstlisting}
\end{minipage}
\subcaption{}
\label{fig:ip_code}
\end{minipage}
\begin{minipage}[t]{0.48\textwidth}
\centering
\begin{tikzpicture}[
  scale=0.7,
  font=\small,
  cell/.style={
    draw=black,
    minimum width=5mm,
    minimum height=0.5cm,
    inner sep=0pt,
  },
]
\matrix (top) [matrix of nodes,
  nodes in empty cells,
  row sep=0pt,
  column sep=0pt,
  nodes={cell},
] {
 & & & & & & \\
};
\node[above=3pt] at (top-1-1.north west) {0};
\node[above=3pt] at (top-1-1.north east) {4};
\node[above=3pt] at (top-1-2.north east) {8};
\node[above=3pt] at (top-1-3.north east) {12};
\node[above=3pt] at (top-1-4.north east) {16};
\node[above=3pt] at (top-1-5.north east) {20};
\node[above=3pt] at (top-1-6.north east) {24};
\draw[<->,line width=1pt] ($(top-1-1.south east)+(0,-0.08cm)$) to[bend right=60] ($(top-1-2.south east)+(0,-0.08cm)$);
\node[below=9pt] at ($(top-1-1.south west)!0.5!(top-1-7.south east)$) {$\unifyCells((\node, [4,4]), (\node, [8,8]))$};

\end{tikzpicture}
\subcaption{}
\label{fig:unify_ip_before}
\end{minipage}
\hspace{0.5\fill}
\begin{minipage}[t]{0.48\textwidth}
\centering
\begin{tikzpicture}[
  scale=0.7,
  font=\small,
  cell/.style={
    draw=black,
    minimum width=5mm,
    minimum height=0.5cm,
    inner sep=0pt,
  },
  chunk/.style={pattern=north east lines, pattern color=purple},
  baseline={([yshift=-1.64cm]current bounding box.north)},
]

\matrix (top) [matrix of nodes,
  nodes in empty cells,
  row sep=0pt,
  column sep=0pt,
  nodes={cell},
  column 2/.style={nodes={minimum width=2cm}}, 
] {
   & |[chunk, minimum width=10mm]| & & & & \\
};
\node[above=3pt] at (top-1-1.north west) {0};
\node[above=1pt] at (top-1-2.north) {[4,8]};
\node[above=3pt] at (top-1-2.north east) {12};
\node[above=3pt] at (top-1-3.north east) {16};
\node[above=3pt] at (top-1-4.north east) {20};
\node[above=3pt] at (top-1-5.north east) {24};
\end{tikzpicture}
\subcaption{}
\label{fig:unify_ip_after}
\end{minipage}
\caption{Effect of $\unifyCells$ on the same node: (a) code; (b) before; (c) after.}
\label{fig:ipcode-unify}
\end{figure} Unfortunately, unification does not always preserve field sensitivity. In DSA, cells carry constant offsets, and unification must merge the two cells into one cell with one offset. When they are in distinct nodes ($\node_1 \neq \node_2$), DSA shifts one node so that the two offsets match and then merges the two nodes; e.g., unifying $(\node_1,0)$ with $(\node_2,4)$ shifts every cell of $\node_1$ by the relative offset $\delta=4$. When they are in the same node ($\node_1 = \node_2$) at different offsets, as for the two targets of \code{addr} in \cref{fig:ip_code} (explained later), this is not allowed because a node cannot be shifted relative to itself. DSA then collapses the node and loses field sensitivity. \newdsa avoids this by using interval offsets: unifying two same-node cells creates a merged cell with offset $\absof_1 \offsetjoin \absof_2$ (e.g., $[0,0]\offsetjoin[4,4]=[0,4]$), merges only cells overlapping that interval, and leaves the others unchanged. The rest of this section presents these two unification cases.

\begin{figure}[t]
\centering
\footnotesize
\begin{minipage}[t]{0.5\linewidth}
\begin{algorithmic}
\MyFunction{$\unifyCells$}{$\cell_1, \cell_2$} \Comment{$\omin{\offsetfunc{\cell_1}} \le \omin{\offsetfunc{\cell_2}}$}
  \If{$\nodefunc{\cell_1} = \nodefunc{\cell_2}$} \Comment{$\unifyCellsSame$}
    \State $\absof := \offsetfunc{\cell_1} \offsetjoin \offsetfunc{\cell_2}$
    \State $S := \{\cell \in \cellset_{\nodefunc{\cell_1}} \mid \offsetfunc{\cell} \offsetmeet \absof \neq \offsetbot\}$
    \State $\mergeCells(S, \absof)$
  \Else \Comment{$\unifyCellsOther$}
    \State $\delta := \omin{\offsetfunc{\cell_2}} - \omin{\offsetfunc{\cell_1}}$
    \State $\unifyNodes(\nodefunc{\cell_1}, \nodefunc{\cell_2}, \delta)$
  \EndIf
\EndMyFunction
\end{algorithmic}
\end{minipage}\hfill
\begin{minipage}[t]{0.465\linewidth}
\begin{algorithmic}
\MyFunction{$\unifyNodes$}{$\node_1, \node_2, \delta$}
  \ForAll{$\cell_i \in \cellset_{\node_1}$}
    \State $\absof_i := \offsetfunc{\cell_i} \offsetadd [\delta, \delta]$
    \State $S := \{\cell \in \cellset_{\node_2} \mid \offsetfunc{\cell} \offsetmeet \absof_i \neq \offsetbot\}$
    \State $\cell_m := \mergeOrNew(S, \node_2, \absof_i)$
    \State $\redirectEdges(\cell_i, \cell_m)$
  \EndForAll
  \State $\nodeset := \nodeset \setminus \{\node_1\}$; $\cellset := \cellset \setminus \cellset_{\node_1}$
\EndMyFunction
\end{algorithmic}
\end{minipage}

\vspace{\baselineskip}

\begin{minipage}[t]{0.5\linewidth}
\begin{algorithmic}
\MyFunction{$\mergeCells$}{$S, \absof$} \Comment{$S = [\cell_1, \ldots, \cell_k]$}
  \State $\cell_m := (\nodefunc{\cell_1}, (\sqcup_{\cell \in S} \offsetfunc{\cell}) \offsetjoin \absof)$
  \State $\cellset := (\cellset \setminus S) \cup \{\cell_m\}$
  \ForAll{$i \in [1, k]$}
    \State $\redirectEdges(\cell_i, \cell_m)$
  \EndForAll
  \Return $\cell_m$
\EndMyFunction
\end{algorithmic}
\end{minipage}\hfill
\begin{minipage}[t]{0.465\linewidth}
\begin{algorithmic}
\MyFunction{$\mergeOrNew$}{$S, \node, \absof$}
  \If{$S = \emptyset$} \Comment{$\mergeOrNew$-\textsc{New}}
    \State $\cellset := \cellset \cup \{(\node, \absof)\}$
    \Return $(\node, \absof)$
  \Else \Comment{$\mergeOrNew$-\textsc{Merge}}
    \Return $\mergeCells(S, \absof)$
  \EndIf
\EndMyFunction
\end{algorithmic}
\end{minipage}
\caption{Unification procedures (Part I). They update the graph $\graph$ in place.}
\label{fig:unification}
\end{figure}

\begin{figure}[t]
\centering
\footnotesize
\begin{minipage}[t]{0.56\linewidth}
\begin{algorithmic}
\MyFunction{$\redirectEdges$}{$\cell_1, \cell_2$}
  \State $\graph := \graph[\cell_2/\cell_1]_{\mathrm{value}}$
  \If{$\cell_1, \cell_2 \in \dom(\edgeset)$} \Comment{$\redirectEdges$-\textsc{Rec}}
    \State $\unifyCells(\edgeset(\cell_1), \edgeset(\cell_2))$
  \Else\ \textbf{skip} \Comment{$\redirectEdges$-\textsc{Easy}}
  \EndIf
  \State $\edgeset := \edgeset[\cell_2/\cell_1]_{\mathrm{key}}$
\EndMyFunction
\end{algorithmic}
\end{minipage}\hfill
\begin{minipage}[t]{0.42\linewidth}
\begin{algorithmic}
\MyFunction{$\lookUp$}{$m, k$}
  \If{$k \in \dom(m)$} \Comment{$\lookUp$-1}
    \Return $m(k)$
  \EndIf
  \State $\node \notin \nodeset$; $\nodeset := \nodeset \cup \{\node\}$ \Comment{$\lookUp$-0}
  \State $\cellset := \cellset \cup \{(\node, [0,0])\}$
  \State $m := m \cup \{k \mapsto (\node, [0,0])\}$
  \Return $(\node, [0,0])$
\EndMyFunction
\end{algorithmic}
\end{minipage}
\caption{Unification procedures (Part II).}
\label{fig:unification-helpers}
\end{figure}

The unification procedures of \newdsa are given in \cref{fig:unification,fig:unification-helpers}. The arguments of $\unifyCells$ are cells $\cell_1, \cell_2 \in \cellset$ of $\graph$ with $\omin{\offsetfunc{\cell_1}} \leq \omin{\offsetfunc{\cell_2}}$ (otherwise, they are swapped). The operation unifies $\node_1$ into $\node_2$, redirects incoming edges to $\node_2$, and recursively unifies outgoing edges. For brevity, we omit the case $\unifyCells(\cell, \cell)$, which leaves $\graph$ unchanged. We also omit forwarding: after a cell such as $\cell_2$ in $\redirectEdges$ is merged, later steps use the cell it was merged into.

For same-node unification ($\unifyCellsSame$), \newdsa computes the joined interval $\absof$, merges all and only cells (in a set $S$) overlapping $\absof$ into one (denoted as $\cell_m$) using $\mergeCells$ (shown in \cref{fig:unification}), and leaves the rest of the cells of the given node unchanged. During $\mergeCells$, it is necessary to redirect edges for each cell in $S$ to $\cell_m$ following $\redirectEdges$ (explained later). During this process, if we compute $\absof = \offsettop$, this indicates that the current node is collapsed.

We illustrate $\unifyCells$ on a single DSA node using the example~\cite{curl-if2ip} in \cref{fig:ip_code}. At \cref{cline:ipaddr1,cline:ipaddr2}, \code{addr} points to the raw IP address according to the protocol indicated by \code{af}. Since IPv4 and IPv6 store the address in different sub-structures, the assignment to \code{addr} requires a different offset. We show the DSA node $\node$ representing the object referenced by \code{iface->ifa_addr} in \cref{fig:unify_ip_before}. The node $\node$ includes two distinct cells for the raw IP addresses, $\{(\node, [4, 4]), (\node, [8, 8])\}$. Since \code{addr} cannot point to two cells at once by C.3, $\unifyCells((\node, [4, 4]),\allowbreak (\node, [8, 8]))$ is invoked to merge the two cells into $(\node, [4, 8])$, as shown in \cref{fig:unify_ip_after}.

For unification across distinct nodes ($\unifyCellsOther$), the two nodes are first aligned and then merged by $\unifyNodes$. The alignment uses a relative offset $\delta$, computed as the difference between the minimum (or starting) offsets of the two cells, to shift $\node_1$ for unification compatibility.

In $\unifyNodes$, each cell $\cell_i$ of $\node_1$ is shifted by $\delta$ (denoted as $\absof_i$). To ensure unified results maintain cell disjointness, we collect cells $S$ in $\node_2$ that overlap $\absof_i$, and then call $\mergeOrNew$, which either merges the cells in $S$ ($\mergeOrNew$-\textsc{Merge}) or creates a new one ($\mergeOrNew$-\textsc{New}). If $S = \emptyset$, we create a new cell $\cell_m$ in $\node_2$, add it to $\cellset$, and redirect all edges of $\cell_i$ to $\cell_m$ using $\redirectEdges$ in \cref{fig:unification-helpers}. If $S \neq \emptyset$, we merge all cells in $S$ into a single cell $\cell_m$ by $\mergeCells$ and then redirect the edges of $\cell_i$ to $\cell_m$. Once all cells of $\node_1$ have been redirected, $\node_1$ and its cells are removed from the graph, so each unification merges two nodes into one and keeps the node set finite. This ensures termination. 

The procedure $\redirectEdges$ moves all edges from $\cell_1$ to $\cell_2$ and has two cases. If the given $\cell_1$ and $\cell_2$ both have outgoing edges, we need to recursively unify their points-to cells to maintain the invariant C.3. Once the recursive unification is done, we then update the edge set key to remove $\cell_1$. Otherwise, we replace $\cell_1$ with $\cell_2$ in the graph ($\redirectEdges$-\textsc{Easy}).

\begin{figure}[t]
\begin{minipage}[t]{0.5\textwidth}
\centering
\begin{tikzpicture}[
  scale=0.5,
  x=0.28cm,
  y=0.95cm,
  font=\small,
  >=Stealth,
  mem/.style={draw=black, line width=1pt},
  chunk/.style={pattern=north east lines, pattern color=purple},
  baseline={(current bounding box.center)},
  guideB/.style={black, densely dotted, line width=0.8pt},
  guideP/.style={purple, densely dashed, line width=0.8pt},
  thickBrown/.style={brown!70!black, line width=2pt},
  thickGray/.style={gray!70, line width=2pt},
]

\def\H{0.65}

\def\xTop{4*1.5}
\def\yTop{0}
\def\nOneBot{-0.5*\H}

\def\xLeft{0}
\def\yBot{-2.2}

\newcommand{\NodeOneFrame}[3]{%
  \draw[mem] (#1,#2) rectangle ++(30,\H);
  \node[left] at (#1,{#2+0.5*\H}) {$#3$};
}

\newcommand{\NodeTwoFrame}[3]{%
  \draw[mem] (#1,#2) rectangle ++(33,\H);
  \node[left] at (#1,{#2+0.5*\H}) {$#3$};
}

\newcommand{\CellBelow}[5]{%
  \fill[chunk] ({#1+#3*1.5},#2) rectangle ({#1+#4*1.5},{#2+\H});
  \draw[mem]  ({#1+#3*1.5},#2) rectangle ({#1+#4*1.5},{#2+\H});
  \node[below=3pt] at ({#1+0.5*(#3*1.5+#4*1.5)},#2) {$#5$};
}

\newcommand{\TopCell}[7][0]{%
  \fill[chunk] ({#2+#4*1.5},#3) rectangle ({#2+#5*1.5},{#3+\H});
  \draw[mem]  ({#2+#4*1.5},#3) rectangle ({#2+#5*1.5},{#3+\H});
  \node[black, above=2pt]  at ({#2+#1+0.53*(#4*1.5+#5*1.5)},{#3+\H}) {$#6$};
  \node[orange, above=14pt]           at ({#2+#1+0.55*(#4*1.5+#5*1.5)},{#3+\H}) {$#7$};
}

\NodeOneFrame{\xTop}{\yTop}{\node_1}

\TopCell[-0.6]{\xTop}{\yTop}{0}{6}{[0,24]}{[16,40]}
\fill[chunk] ({\xTop+7*1.5},\yTop) rectangle ({\xTop+7*1.5},{\yTop+\H});
\draw[mem]  ({\xTop+7*1.5},\yTop) rectangle ({\xTop+7*1.5},{\yTop+\H});
\node[black, above=4pt]  at ({\xTop+0.49*(7*1.5+7*1.5)},{\yTop+\H}) {$28$};
\node[orange, above=16pt] at ({\xTop+0.48*(7*1.5+7*1.5)},{\yTop+\H}) {$44$};
\TopCell[1.6]{\xTop}{\yTop}{8}{10}{[32,40]}{[48,56]}
\TopCell{\xTop}{\yTop}{12}{20}{[48,80]}{[64,96]}

\draw[orange, densely dotted, |->] 
    (\xLeft,{\nOneBot+0.5*\H}) -- (\xTop,{\nOneBot+0.5*\H})
    node[midway, below=1pt]  {$\delta = 16$};

\NodeTwoFrame{\xLeft}{\yBot}{\node_2}

\CellBelow{\xLeft}{\yBot}{0}{2}{[0,8]}
\CellBelow{\xLeft}{\yBot}{4}{8}{[16,32]}
\CellBelow{\xLeft}{\yBot}{12}{18}{[48,72]}
\CellBelow{\xLeft}{\yBot}{20}{22}{[80,88]}

\coordinate (Utop) at ({\xTop+9*1.5},{\yTop});
\coordinate (Ubot) at ({\xLeft+13*1.5},{\yBot+\H});
\draw[<->,line width=1pt] (Utop) -- (Ubot);
\node[font=\scriptsize,right=5pt] at ($(Utop)!0.5!(Ubot)$) {$\unifyCells(\cell_1, \cell_2)$};
\end{tikzpicture}
\subcaption{}
\label{fig:unify_diff_nodes_before}
\end{minipage}%
\hspace{0.5\fill}%
\begin{minipage}[t]{0.5\textwidth}
\centering
\begin{tikzpicture}[
  scale=0.5,
  x=0.28cm,
  y=0.95cm,
  font=\small,
  >=Stealth,
  mem/.style={draw=black, line width=1pt},
  chunk/.style={pattern=north east lines, pattern color=purple},
  baseline={(current bounding box.center)},
  guideB/.style={black, densely dotted, line width=0.8pt},
  guideP/.style={purple, densely dashed, line width=0.8pt},
  thickBrown/.style={brown!70!black, line width=2pt},
  thickGray/.style={gray!70, line width=2pt},
  baseline={([yshift=.52cm]current bounding box.north)},
]
\def\H{0.65}

\def\nOneBot{-0.5*\H}

\def\xLeft{0}
\def\yBot{0}

\def\xRight{31}
\def\yRight{-2.2}

\newcommand{\NodeFrame}[3]{%
  \draw[mem] (#1,#2) rectangle ++(24*1.5,\H);
  \node[left] at (#1,{#2+0.5*\H}) {$#3$};
}

\newcommand{\CellBelow}[5]{%
  \fill[chunk] ({#1+#3*1.5},#2) rectangle ({#1+#4*1.5},{#2+\H});
  \draw[mem]  ({#1+#3*1.5},#2) rectangle ({#1+#4*1.5},{#2+\H});
  \node[below=3pt] at ({#1+0.5*(#3*1.5+#4*1.5)},#2) {$#5$};
}

\NodeFrame{\xLeft}{\yBot}{\node_2}

\CellBelow{\xLeft}{\yBot}{0}{2}{[0,8]}
\CellBelow{\xLeft}{\yBot}{4}{10}{[16,40]}
\fill[chunk] ({\xLeft+11*1.5},\yBot) rectangle ({\xLeft+11*1.5},{\yBot+\H});
\draw[mem]  ({\xLeft+11*1.5},\yBot) rectangle ({\xLeft+11*1.5},{\yBot+\H});
\node[below=4pt] at ({\xLeft+0.5*(11*1.5+11*1.5)},{\yBot}) {$44$};
\CellBelow{\xLeft}{\yBot}{12}{24}{[48,96]}

\end{tikzpicture}
\subcaption{}
\label{fig:unify_diff_nodes_after}
\end{minipage}
\caption{Effect of $\unifyCells$ on different nodes: (a) before; (b) after.}
\label{fig:unify_diff_nodes}
\end{figure} We provide an example of unification across distinct nodes in \cref{fig:unify_diff_nodes_before}, where $\cell_1 = (\node_1, [32, 40])$ and $\cell_2 = (\node_2, [48, 72])$. When $\unifyCells(\cell_1, \cell_2)$ is called, the $\unifyCellsOther$ case computes a relative offset $\delta = \omin{\offsetfunc{\cell_2}} - \omin{\offsetfunc{\cell_1}} = 16$ and shifts $\node_1$ by $\delta$ in $\unifyNodes$. Each shifted cell of $\node_1$ is then unified with overlapping cells in $\node_2$. If no overlap exists, such as $(\node_1,[28,28])$ shifted to $[44,44]$, a new cell covering the range is created. If one overlap exists, such as $(\node_1,[0,24])$ shifted to $[16,40]$ and $(\node_2,[16,32])$, a new cell $(\node_2,[16,40])$ is created to cover both. If multiple overlaps exist, such as $(\node_1,[48,80])$ shifted to $[64,96]$, which overlaps $(\node_2,[48,72])$ and $(\node_2,[80,88])$, $\mergeCells$ computes a cell $(\node_2, [48,96])$. We apply $\redirectEdges$ when cells are merged. Finally, cells of $\node_2$ with no overlap, such as $(\node_2, [0,8])$, are preserved. The resulting $\node_2$ is shown in \cref{fig:unify_diff_nodes_after}.

\begin{figure}[t]
\centering
\begin{minipage}[t]{0.48\linewidth}
\begin{algorithmic}
\MyFunction{$\dsatranon{\pointerp = \makeref(\inti)}$}{$\graph$}
  \State $\freshnode \notin \nodeset$; $\nodeset := \nodeset \cup \{\freshnode\}$
  \State $\cellset := \cellset \cup \{(\freshnode, [0,0])\}$
  \State $\unifyCells(\lookUp(\envmap, \pointerp), (\freshnode, [0,0]))$
\EndMyFunction
\end{algorithmic}
\end{minipage}\hfill
\begin{minipage}[t]{0.48\linewidth}
\begin{algorithmic}
\MyFunction{$\dsatranon{\varx = *\pointerp \,/\, *\pointerp = \varx}$}{$\graph$} \Comment{$\varx \in \varptrset$}
  \State $\cell_\pointerp := \lookUp(\envmap, \pointerp)$
  \State $\cell_d := \lookUp(\edgeset, \cell_\pointerp)$
  \State $\unifyCells(\cell_d, \lookUp(\envmap, \varx))$
\EndMyFunction
\end{algorithmic}
\end{minipage}

\vspace{\baselineskip}

\begin{minipage}[t]{0.48\linewidth}
\begin{algorithmic}
\MyFunction{$\dsatranon{\pointerp = \pointerq + \of}$}{$\graph$}
  \State $(\node, \absof_\pointerq) := \lookUp(\envmap, \pointerq)$
  \State $\absof := \absof_\pointerq \offsetadd [\of, \of]$
  \State $S := \{\cell \in \cellset_\node \mid \offsetfunc{\cell} \offsetmeet \absof \neq \offsetbot\}$
  \State $\cell_m := \mergeOrNew(S, \node, \absof)$
  \State $\unifyCells(\lookUp(\envmap, \pointerp), \cell_m)$
\EndMyFunction
\end{algorithmic}
\end{minipage}\hfill
\begin{minipage}[t]{0.48\linewidth}
\begin{algorithmic}
\MyFunction{$\dsatranon{\pointerp = \pointerq + (\inti \times \elementsize)}$}{$\graph$}
  \State $(\node, \absof_\pointerq) := \lookUp(\envmap, \pointerq)$
  \State $\absof := \absof_\pointerq \offsetadd [0, +\infty]$
  \State $S := \{\cell \in \cellset_\node \mid \offsetfunc{\cell} \offsetmeet \absof \neq \offsetbot\}$
  \State $\cell_m := \mergeOrNew(S, \node, \absof)$
  \State $\unifyCells(\lookUp(\envmap, \pointerp), \cell_m)$
\EndMyFunction
\end{algorithmic}
\end{minipage}
\caption{The transfer functions of \newdsa{}: each updates $\graph$ in place and returns it.}
\label{fig:dsa-trans}
\end{figure}

Our transfer functions $\dsatran: \graphset \to \graphset$ for \newdsa are shown in \cref{fig:dsa-trans}. The analysis builds a DSG by evaluating pointer-manipulating statements. Since it is flow-insensitive, when a statement queries a points-to relation, the analysis either reuses an existing one ($\lookUp$-1) or creates a new one ($\lookUp$-0).

Specifically, for allocation, the analysis creates a fresh node $\freshnode$ for the allocated object and adds a points-to edge $\pointerp \mapsto (\freshnode, [0,0])$, using unification if needed. 
For struct-field address computation, it obtains the cell of $\pointerq$'s node $\node$ covering the cumulative offset $\absof = \absof_\pointerq \offsetadd [\of,\of]$ with $\mergeOrNew$, which creates a new cell only if no cell overlaps $\absof$ and otherwise merges the overlapping cells, as C.1 requires. The analysis then unifies the cell of $\pointerp$ with this cell. For example, after \code{b->buf[i]}, \code{b->buf[4]} gets the interval cell $(\rb,[12,+\infty])$, not a new cell $(\rb,[28,28])$.
For array pointer arithmetic with symbolic index $\inti$, the analysis approximates the possible index range and obtains a cell covering all reachable array elements. Since \taintir{} restricts pointer arithmetic to non-negative offsets (\cref{sec:concrete_lan_sem}), we approximate the range of $\inti$ as $[0, +\infty]$. This range can be refined using precomputed value information (e.g., from value-set analysis) or when the array size is statically known. 
For a memory load, if the loaded value assigned to $\varx$ is a pointer, the analysis looks up the cell for $\varx$ and unifies it with the cell reached by dereferencing $\pointerp$. A store is handled similarly. A primitive statement without a rule in \cref{fig:dsa-trans} leaves $\graph$ unchanged. For a non-primitive statement $\statements$, $\dsatran$ applies the transfer functions of the primitive statements in $\statements$ repeatedly until it reaches a fixpoint. The range of a symbolic array index $\inti$ is already $[0,+\infty]$, but pointer arithmetic in a loop, such as $\pointerp = \pointerp + 4$, enlarges an interval in every iteration. We therefore use the standard interval widening~\cite{CousotIntervals}, which replaces an unstable upper bound by $+\infty$ and an unstable lower bound by $0$, the smallest offset: $[a,b] \mathbin{\nabla} [c,d] = [\,c<a \mathrel{?} 0 : a,\ d>b \mathrel{?} {+\infty} : b\,]$. We apply it only where pointer arithmetic grows a cell, and only after one join for a singleton $[a,a]$ (see \appref{proof:dsa-soundness}).

For the code in \cref{fig:redis_code}, \newdsa eventually constructs the graph at the top of \cref{fig:flow_redis} with cells: $\cell_{\mathit{id}}=(\rb, [0,0])$, $\cell_{\mathit{size}}=(\rb, [4,4])$, $\cell_{\mathit{used}}=(\rb, [8, 8])$, and $\cell_{\mathit{buf[]}}=(\rb, [12, +\infty])$. 
At \cref{cline:rdballoc}, the analysis creates the node $\rb$ and maps \code{b} to $(\rb, [0,0])$. At \cref{cline:rdidassign,cline:rdszassign,cline:rdusedinit}, it establishes points-to relations for the intermediate pointers \code{bid}, \code{bsize}, and \code{bused} to cells $\cell_{\mathit{id}}$, $\cell_{\mathit{size}}$, and $\cell_{\mathit{used}}$, respectively. The array access at \cref{cline:rdbufassign} then constructs $\cell_{\mathit{buf[]}}$, referenced by \code{bbufi}. 

\begin{theorem}[Correctness of $\dsatran$]
\label{thm:dsa-soundness}
\[\forall \statements \in \statement,\ \forall \graph \in \graphset ~.~
\concretetran(\concretegraph(\graph)) \subseteq \concretegraph(\dsatran(\graph))
\]
\end{theorem}
\noindent The proof is given in \appref{proof:dsa-soundness}.

In summary, we have shown how the analysis constructs the points-to graph for a single function. 
While \newdsa uses intervals to abstract offsets, the approach can be instantiated with other non-relational domains, such as constant sets or reduced products of intervals and congruences. DSA also propagates graph information between callers and callees to support context sensitivity. This mechanism is orthogonal to our extension, and \newdsa retains it~\cite{DBLP:conf/pldi/LattnerA05,DBLP:conf/pldi/LattnerLA07}. As discussed in \cref{sec:overview}, the precision of a taint analysis depends on the field sensitivity of its PTA, which \newdsa{} is designed to preserve. Next, we present the DFA that propagates taint over the graphs computed by \newdsa{}.
\section{Taint Semantics}
\label{sec:taint_sem}
This section presents a data-flow analysis (DFA) built on the \newdsa memory abstraction of \cref{sec:dsa_o} to track propagation through explicit data dependencies. We track only such \emph{explicit} flows, which is the standard scope for a taint analysis built as a client of a points-to analysis. \emph{Implicit} flows from control dependencies are left to future work. We first define the rules governing taint flow across variables and memory locations, and then formalize the analysis as a reduced product of a memory domain and a taint domain. The numerical invariants of the memory domain let the analysis prune taint on provably infeasible paths.

We extend the concrete state $\state = (\cenv,\heap)$ to a tainted state $\taintstate = (\cenv,\heap,\taintMap) \in \taintstateset$, where $\taintMap \in \powerset(\locs)$ records tainted locations and $\locs \Def= \varset \cup \ptrvalset$ includes program variables and memory addresses. This set representation is equivalent to a map $\locs \to \taintLattice$, where $\taintLattice=\{\clean,\tainted\}$ is ordered by $\clean \taintleq \tainted$ with $\clean \taintjoin \tainted = \tainted$. Since $\taintLattice$ has only two elements, we store $\loc \in \taintMap$ when $\loc$ is tainted. The set generalizes to a map from locations to any finite lattice.

To track taint flows through data dependencies, we define a propagation rule that updates the set of tainted locations. The rule describes a taint flow from source locations $\srcset$ to target locations $\tgtset$, marking $\tgtset$ as tainted if and only if some source in $\srcset$ is tainted. Formally, $\taintMap' := \propshort{\srcset}{\tgtset} \Def= \taintMap \setminus \tgtset \cup \{\loc \in \tgtset \mid \srcset \cap \taintMap \neq \emptyset\}$.

\cref{fig:trans} \newwork{highlights} the propagation rules of the transfer function $\taintctran: \taintstateset \to \taintstateset$. When \code{input} introduces a taint source, we add the tainted variable directly to $\taintMap$. A sanitizer \code{sanitize(x)} removes \code{x} from $\taintMap$. A sink \code{output(x)} does not change the state but leaks tainted data if $\varx \in \taintMap$. For assignment statements, the propagation rule follows data dependencies between variables. For example, for \code{x = input(); y = x + 1; z = 3}, the rule adds \code{x} to the taint set via $\taintMap \cup \{\varx\}$. Subsequently, $\propshort{\varx}{\mathit{y}}$ (short for $\propshort{\{\varx\}}{\{\mathit{y}\}}$) adds \code{y} since $\varx \in \taintMap$, while the rule for \code{z} does not add \code{z}, following $\propshort{\{\}}{\mathit{z}}$.

For allocation, our analysis does not introduce taint, even if the allocation size is tainted. More conservative analyses may propagate taint from allocation sizes to the resulting pointer, $\propshort{\inti}{\pointerp}$ (e.g., \cite{DBLP:conf/uss/0001LJHAY23}), or to the entire allocated object, $\propshort{\inti}{\{\pointerp, \ptr{\baddr}{0},\ldots, \ptr{\baddr}{n}\}}$ (e.g., \cite{DBLP:journals/tc/DoudalisCVPO12}). For pointer arithmetic, we propagate taint only between pointers and ignore the computed offset. For memory load and store operations, taint propagates between the variable and the memory location of the stored content. For instance, given \code{p = alloc(4); x = input(); *p = x}, the analysis propagates taint from \code{x} to the memory location pointed to by \code{p}. Similarly, \code{x = *p; sanitize(x); *p = x} removes taint from the memory location pointed to by \code{p}.
Overall, we write $\taintctran :\taintdom \to \taintdom$ for the collecting taint semantics induced by the statement-level transfer function, where $\taintdom \Def = \powerset(\taintstateset)$.

Our DFA is combined with a memory-aware numeric analysis that tracks numerical invariants. Let $\numdom$ be a numerical domain, instantiated, for example, as Intervals~\cite{CousotIntervals}, Octagons~\cite{DBLP:conf/wcre/Mine01}, or Polyhedra~\cite{DBLP:conf/popl/CousotH78}.
Formally, we define the abstract domain $\abstaintdom \Def= \absmemdom \times \abssetdom$ as a reduced product of a memory domain $\absmemdom$ and a taint domain $\abssetdom$, with a reduction operator $\mathcal{R} : \abstaintdom \rightarrow \abstaintdom$ for normalization. The memory domain $\absmemdom$ uses $\numdom$ to capture relational or non-relational properties over abstract memory locations $\abslocs$. The taint domain is defined as $\abssetdom \Def= \powerset(\abslocs)$, where $\absloc \in \abstaintstate \in \abssetdom$ means that $\absloc$ may be tainted. We use this set representation instead of a map from locations to elements of the lattice $\taintLattice$.

Analysis precision is determined by the granularity of abstract locations $\abslocs$, which form the dimensions of our product domain. Each dimension carries two abstract values: a numerical bound and a taint element, both under the same memory abstraction. We instantiate this abstraction with \newdsa, representing memory locations as variables and memory cells. After \newdsa constructs its DSGs $\graph$, we define $\abslocs \Def= \varset \cup \bigcup_{\graph} \graph.\cellset$. Following the summarization approach in~\cite{DBLP:conf/pldi/ChaseWZ90}, $\absmemdom$ treats memory objects as summary objects (DSA nodes) and models reads and writes with specialized operations.
We omit memory-domain design and implementation details because they are orthogonal to taint abstraction, and assume $\absmemdom$ provides $\membot$ (bottom), $\memleq$ (partial order), and $\memjoin$ (join).

The abstract state $\absstate \in \abstaintdom$ consists of a memory state $\absmemstate \in \absmemdom$ and a taint state $\abstaintstate \in \abssetdom$. 
The taint state keeps a set of abstract locations (variables or memory cells) that are tainted, and the propagation function $\abstaintstate_1 := \abspropshort{\abssrcset}{\abstgtset} \Def= \abstaintstate \setminus \abskills(\abstgtset) \cup \{\absloc \in \abstgtset \mid \abssrcset \cap \abstaintstate \neq \emptyset\}$ soundly abstracts the concrete rule. $\abskills$ filters a location set and keeps only locations whose taint can be safely cleared. We allow killing taint on variables, but not on memory cells. Since each cell summarizes multiple object fields, we conservatively keep cells tainted. Formally, we define $\abskills(\abstgtset) = \abstgtset \cap \varset$.

\begin{figure}[t]
\centering
\begin{tabular*}{\linewidth}{@{}r@{\;}l@{\extracolsep{\fill}}r@{}}
$\dfatranon{\varx = *\pointerp}((\absmemstate, \abstaintstate)) =$ & $(\memtran(\absmemstate),\, \abspropshort{\{\cell\}}{\{\varx\}})$ & $\cell = \graph.\envmap(\pointerp)$ \\
$\dfatranon{*\pointerp = \varx}((\absmemstate, \abstaintstate)) =$ & $(\memtran(\absmemstate),\, \abspropshort{\{\varx\}}{\{\cell\}})$ & $\cell = \graph.\envmap(\pointerp)$
\end{tabular*}
\caption{Abstract transfer functions $\dfatran$ for taint analysis.}
\label{fig:dfa-trans}
\end{figure} 
The abstract transfer function $\dfatran: \abstaintdom \to \abstaintdom$ combines the transfer functions of its underlying sub-domains. We assume $\absmemdom$ provides a sound transfer function $\memtran: \absmemdom \to \absmemdom$ for reasoning about numerical properties of abstract locations $\abslocs$. Initially, the abstract state maps all locations to $\clean$ ($\abstaintstate = \emptyset$) and all numerical elements to $\top$. Most abstract taint rules, including those for integer/pointer arithmetic, external input, and sanitizers, use the same source and target location sets as the concrete semantics, so we only show load and store in \cref{fig:dfa-trans}. For these memory operations, the analysis follows the precomputed DSA graph $\graph$ to identify the target cell $\cell$ and propagates taint between that cell and the variable to soundly approximate the data flow of the access. For \code{sanitize(x)}, the abstract rule removes $\abskills(\{\varx\}) = \{\varx\}$ from $\abstaintstate$. Taint on a cell is not killed, so the cell pointed to by \code{p} stays tainted after \code{x = *p; sanitize(x); *p = x}. At \code{output(x)}, the analysis reports a warning if $\varx \in \abstaintstate$.

The reduction $\mathcal{R}$ uses the memory domain to prune taint facts in unreachable states: $\mathcal{R}((\absmemstate, \abstaintstate)) = (\absmemstate, \emptyset)$ if $\absmemstate = \membot$, and $(\absmemstate, \abstaintstate)$ otherwise.

We assume the memory domain has a concretization function $\concretemem : \absmemdom \to \concretdom$. A concrete state $(\cenv, \heap, \taintMap)$ is represented by an abstract state $(\absmemstate, \abstaintstate)$ when its memory is represented by $\absmemstate$ and simulated by $\graph$ through some $\mu$, and every tainted location is covered by $\abstaintstate$: a tainted variable is in $\abstaintstate$, and a tainted address $\addrval$ has its cell $\mu(\addrval)$ in $\abstaintstate$. Formally, for $\absstate = (\absmemstate, \abstaintstate)$, the concretization $\concretetaint : \abstaintdom \to \taintdom$ is
\[
\begin{aligned}
\concretetaint(\absstate) \Def= \{(\cenv,\heap,\taintMap) \mid{} & (\cenv,\heap) \in \concretemem(\absmemstate) \land \exists \mu.\, (\cenv,\heap) \models_\mu \graph \\
& \land \taintMap \subseteq (\abstaintstate \cap \varset) \cup \{\addrval \in \ptrvalset \mid \mu(\addrval) \in \abstaintstate\}\}.
\end{aligned}
\]

\begin{theorem}[Correctness of $\dfatran$]
\label{thm:dfa-soundness}
\[
\forall \absstate \in \abstaintdom, \taintctran(\concretetaint(\absstate)) \subseteq \concretetaint(\dfatran(\absstate))
\]
\end{theorem}
\noindent The theorem states that the abstract transfer function $\dfatran$ over-approximates the collecting taint semantics $\taintctran$ (\cref{fig:trans} with the highlighted parts). The proof is given in \appref{proof:taint-soundness}. \section{Implementation}
\label{sec:impl}
We implement our taint analysis on top of the \seahorn~\cite{DBLP:conf/tacas/GurfinkelKN15} verification framework, targeting LLVM IR~\cite{DBLP:conf/cgo/LattnerA04}.
The implementation has two stages. The first stage constructs a memory abstraction with \newdsa{}, which we build on top of \seadsa~\cite{DBLP:conf/sas/GurfinkelN17}, a context-, field-, and array-sensitive pointer analysis.

The second stage performs data-flow analysis using \crab~\cite{DBLP:conf/vstte/GurfinkelN21}, an abstract-interpretation library. \crab{} has a language-independent IR (\crabir), whose statements are annotated with explicit memory cells inferred by \newdsa. This lets our taint analysis track taint across memory locations. We use the region domain~\cite{DBLP:conf/vstte/GurfinkelN21} of \crab{} as the memory domain. It relies on object summarization~\cite{DBLP:conf/pldi/ChaseWZ90} to represent multiple concrete memory objects as a single abstract object. On top of this implementation, we add the taint domain of \cref{sec:taint_sem} with its reduction, which discards taint in states that the numerical invariants prove unreachable.

\begin{figure}[t]
\centering
\begin{tikzpicture}[
  font=\scriptsize,
  box/.style={draw, rounded corners, thick, minimum width=1.9cm, minimum height=0.75cm, inner sep=2pt, align=center},
  codebox/.style={draw=none, inner sep=0pt, align=left},
  arrow/.style={-{Latex[length=2mm]}, thick},
  lab/.style={font=\scriptsize, align=center}
]
\node[box] (clang) at (0,0) {\tool{Clang}};
\node[box, right=0.5cm of clang] (instr) {IR\\Instrumenter};
\node[box, right=0.95cm of instr] (seadsa) {\seadsa};
\node[box, right=1.45cm of seadsa] (crab) {\crab};
\node[right=0.45cm of crab, lab] (report) {taint\\report};
\draw[arrow] (clang) -- node[above, lab] {.bc} (instr);
\draw[arrow] (instr) -- node[above, lab] {tt.bc} (seadsa);
\draw[arrow] (seadsa) -- node[above, lab] {tt.bc, DSG} (crab);
\draw[arrow] (crab) -- (report);
\node[codebox, anchor=north west] (code) at ($(clang.south west)+(0,-0.5)$) {
\begin{minipage}{0.39\textwidth}
\begin{lstlisting}[language=C, style=mystyle, basicstyle=\ttfamily\tiny, aboveskip=0pt, belowskip=0pt, xleftmargin=8.41pt]
void main() {
  int i; unsigned u;
  char c; bool b;
  scanf("%d %u %c %d", &i, &u, &c, (int*)&b);
  printf("Results are %d + %u\n", i, u);
}
\end{lstlisting}
\end{minipage}
};
\node[codebox, anchor=north west] (yaml1) at ($(seadsa.south west)+(0,-0.5)$) {
\begin{minipage}{0.2\textwidth}
\begin{lstlisting}[language=yaml, basicstyle=\ttfamily\tiny, aboveskip=0pt, belowskip=0pt]
Propagations:
  - Name: scanf
    DstArgs: [-1]
    VariadicType: Dst
    VariadicIndex: 1
\end{lstlisting}
\end{minipage}
};
\node[codebox, anchor=north west] (yaml2) at ($(yaml1.north east)+(0.1,0)$) {
\begin{minipage}{0.18\textwidth}
\begin{lstlisting}[language=yaml, basicstyle=\ttfamily\tiny, aboveskip=0pt, belowskip=0pt, firstnumber=6]
Sinks:
  - Name: printf
    Args: []
    VariadicIndex: 1
\end{lstlisting}
\end{minipage}
};
\node[codebox, anchor=north west] (yaml3) at ($(yaml2.north east)+(0.1,0)$) {
\begin{minipage}{0.13\textwidth}
\begin{lstlisting}[language=yaml, basicstyle=\ttfamily\tiny, aboveskip=0pt, belowskip=0pt, firstnumber=10]
Filters:
  - Name: bzero
    Args: [0]
\end{lstlisting}
\end{minipage}
};
\draw[arrow] ($(code.north -| clang.south)+(0,0.05)$) -- node[left, lab] {.c} ($(clang.south)+(0,-0.08)$);
\draw[arrow] ($(yaml1.north west)+(0.3,0.05)$) -- node[below left, lab, pos=0.4] {.yaml} ($(instr.south)+(0.3,-0.08)$);
\end{tikzpicture}
\caption{Overview of the taint-analysis workflow.}
\label{fig:taint-workflow}
\end{figure}

\cref{fig:taint-workflow} presents the overall taint-analysis workflow. Following the configuration style of \clangtaint~\cite{clang-taint-analysis-config}, we specify sources, sinks, and propagation rules in YAML rather than hard-coding them in the source code. The input program is compiled to LLVM IR and instrumented according to the taint policy. For each function with a taint policy, the IR instrumenter inserts intrinsics at its call sites to encode that policy in LLVM IR. For source functions, the inserted intrinsics mark the designated arguments and return value as tainted. For example, the \code{Propagations} entry for \code{scanf} marks its variadic output arguments (\code{VariadicIndex: 1}) as tainted. For sink functions, assertions are added to check that the designated sink arguments are untainted. For example, the \code{Sinks} entry for \code{printf} checks its variadic arguments before the call. This instrumentation makes taint behavior explicit in the low-level IR, avoids source-level annotations, and keeps the taint policy configurable and analysis-independent. The policy also lists sanitizers (\code{Filters}), whose designated arguments are clean after the call. Since a cell may summarize multiple objects, the analysis clears a cell after a sanitizer or a store only if the cell represents one concrete object.%
\section{Evaluation}
\label{sec:eval}

\begin{figure}[t]
\centering
\begin{minipage}[c]{0.60\linewidth}
\centering \scriptsize \setlength{\tabcolsep}{1.5pt}
\resizebox{\linewidth}{!}{%
\begin{tabular}{lrrrrrrrr}
\toprule
 & & & & \multicolumn{3}{c}{Warnings} & \multicolumn{2}{c}{Time (s)} \\
\cmidrule(lr){5-7} \cmidrule(lr){8-9}
 & \#Bench. & T/O & \#Checks & S & I & Red. & S & I \\
\midrule
SMALL     & 43  & 0 & 126  & 66  & 62  & 6.1\%  & 6.7    & 6.7    \\
IONC      & 25  & 2 & 149  & 77  & 76  & 1.3\%  & 465.4  & 425.2  \\
CURL      & 20  & 0 & 176  & 80  & 67  & 16.2\% & 8.9    & 9.0    \\
COREUTILS & 109 & 2 & 1785 & 234 & 234 & 0.0\%  & 2183.7 & 2244.7 \\
\midrule
Total     & 197 & 4 & 2236 & 457 & 439 & 3.9\%  & 2664.7 & 2685.6 \\
\bottomrule
\end{tabular}}
\end{minipage}
\hfill
\begin{minipage}[c]{0.38\linewidth}
\centering
\includegraphics[width=1.0\linewidth]{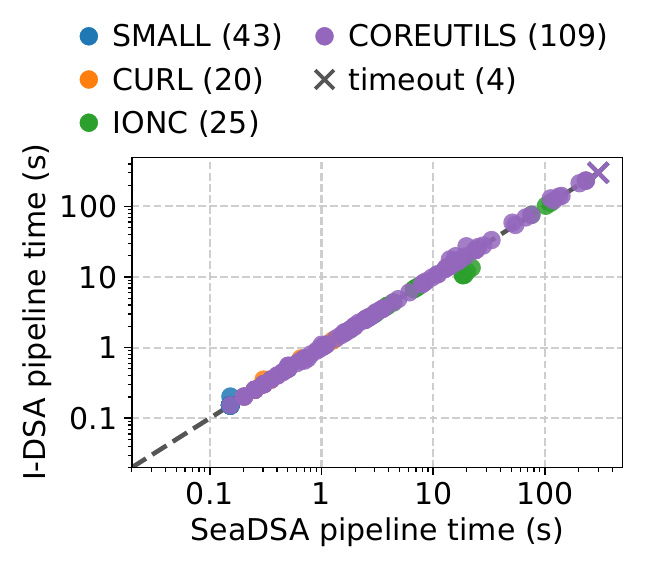}
\end{minipage}
\par
\begin{minipage}[t]{0.60\linewidth}
\subcaption{}
\label{fig:exp1-table}
\end{minipage}
\hfill
\begin{minipage}[t]{0.38\linewidth}
\subcaption{}
\label{fig:exp1-scatter}
\end{minipage}
\caption{\expone{}: (a) results (S = \seadsa{}, I = \newdsa{}); (b) time per benchmark for the whole pipeline.}
\label{fig:exp1}
\end{figure}

\begin{figure}[t]
\centering
\begin{minipage}{\linewidth}
\centering {\scriptsize \setlength{\tabcolsep}{2.5pt}
\begin{tabular}{lrrrrrrrrr}
\toprule
 & & & & \multicolumn{2}{c}{\seadsa{}} & \multicolumn{2}{c}{\newdsa{}} & \multicolumn{2}{c}{\svf{}} \\
\cmidrule(lr){5-6} \cmidrule(lr){7-8} \cmidrule(lr){9-10}
 & \#Bench. & \#Pairs & BasicAA & CI & CS & CI & CS & Andersen & Flow-sens. \\
\midrule
IONC      & 25 & 3\,551\,163 & 46.0\% & 10.0\% & 13.3\% & 10.0\% & 13.3\% & 49.5\% & 49.8\% \\
CURL      & 20 & 503\,879    & 45.0\% & 54.7\% & 70.9\% & 54.7\% & 71.8\% & 75.3\% & 76.8\% \\
COREUTILS & 94 & 9\,293\,743 & 25.0\% & 60.5\% & 69.1\% & 60.5\% & 69.2\% & 72.5\% & 73.0\% \\
\bottomrule
\end{tabular}}
\subcaption{}
\label{tab:exp2}
\end{minipage}
\begin{minipage}{\linewidth}
\centering
\includegraphics[width=0.88\linewidth]{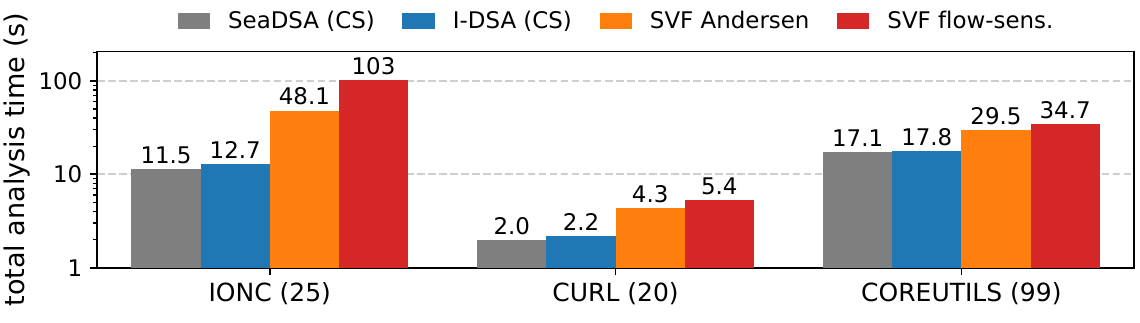}
\subcaption{}
\label{fig:exp2-time}
\end{minipage}
\caption{\exptwo{}: (a) pairs proven never to alias; (b) total analysis time per suite.}
\label{fig:exp2}
\end{figure}

\begin{figure}[t]
\centering
\includegraphics[width=0.42\linewidth]{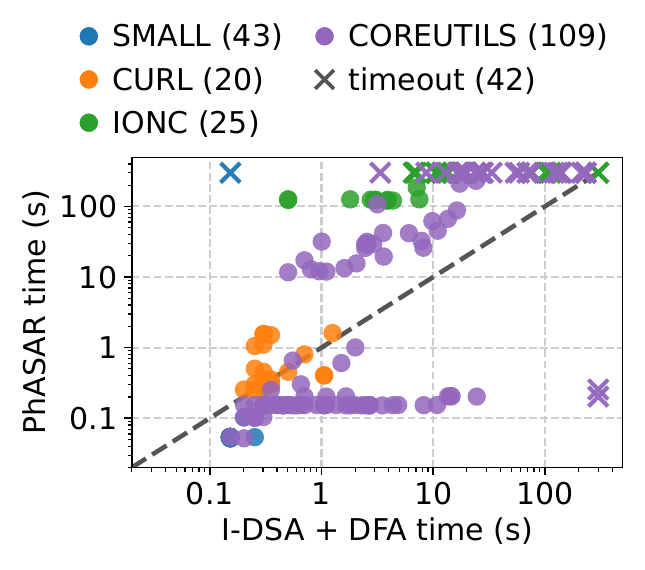}
\caption{\expthree{}: time per benchmark, \newdsa{} + DFA vs.\ \phasar{}.}
\label{fig:exp3-time}
\end{figure}

We evaluate our taint analysis (\newdsa{} + DFA) with three experiments, each against a different existing analysis: the \seadsa{} analysis that \newdsa{} extends, the inclusion-based pointer analysis of \svf{}, and another taint analyzer. Experiment \expone{} in \cref{fig:exp1} compares the baseline configuration, \seadsa{} + DFA, with the proposed configuration, \newdsa{} + DFA, under the same taint policy, so any change in warnings and running time comes from the DSA component alone. As discussed, over-tainting a whole object produces extra warnings, so fewer warnings under the same policy indicate that \newdsa{} preserves field sensitivity. Because \newdsa{} is a pointer analysis, its precision should also hold outside the taint client. Experiment \exptwo{} in \cref{fig:exp2} therefore compares \newdsa{}, which is unification-based, with the inclusion-based pointer analyses of \svf{}~\cite{DBLP:conf/cc/SuiX16} on the same alias queries. Following~\cite{DBLP:conf/sas/DasLFR01,DBLP:conf/pldi/LattnerLA07}, we measure precision as the percentage of pointer pairs that each analysis proves never to alias. Finally, Experiment \expthree{} in \cref{fig:exp3-time} compares our taint analysis end-to-end  with the taint analysis of \phasar{}~\cite{DBLP:conf/tacas/SchubertHB19}, an analysis framework for LLVM IR, under a shared taint policy. Since the two tools propagate taint differently and the real-world benchmarks have no manually verified ground truth, we compare their results per sink and explain the main reasons they disagree. All three experiments report the total running time of each tool.

All experiments were run on an Ubuntu desktop with a $6$-core Intel Core i$5$-$11400$F CPU at $2.60$\,GHz and $96$\,GB RAM, in a container limited to $8$ CPUs and $24$\,GB, running one benchmark at a time. We used a $300$-second timeout for each benchmark: for the whole taint-analysis pipeline in \expone{} and \expthree{}, and for each pointer analysis and each set of alias queries in \exptwo{}. Timeouts are reported separately and excluded from the precision and time comparisons. In \expone{}, both configurations run the same DFA implementation in \crab{} with Intervals as the numerical domain. We report warnings per sink check: a check that runs in several calling contexts counts as a warning if a taint flow is detected in any of them. Every experiment was repeated once to check stability. The reported running times are from the first run.

We evaluated four C benchmark suites with $197$ programs in total. The SMALL suite contains $43$ small programs that we wrote to test core taint-analysis and pointer-analysis functionality. We also included three real-world suites: $25$ benchmarks from IONC, a data-serialization library~\cite{amazon-ion} that converts data between JSON and Amazon Ion; $20$ benchmarks from CURL~\cite{curl}, the library and command-line tool for URL-based data transfer; and $109$ programs from \mbox{COREUTILS}~\cite{gnu_coreutils_project}. For IONC and CURL, each benchmark is a small driver whose \code{main} exercises one API scenario. For COREUTILS, each program is one benchmark with its own \code{main} as the entry point, except the library \code{libstdbuf.so}. A taint policy lists the source, sink, propagation, and sanitizer functions. All benchmarks share a common policy: calls to input functions of the C library and the system (\code{scanf}, \code{read}, \ldots) introduce tainted values, and output functions (\code{printf}, \code{write}, \ldots) are sinks. For the real-world suites, we added each project's own I/O and logging wrappers and, for the drivers, stubs that produce a tainted value or check one. The policy is only for the evaluation and implies no vulnerability in these projects. All benchmarks are compiled with the same compiler flags, and every tool starts its analysis pipeline from the resulting LLVM bitcode.

In \expone{}, \newdsa{} reports fewer warnings than \seadsa{} on SMALL, IONC, and CURL, and the same warnings on COREUTILS (\cref{fig:exp1-table}). The difference is $18$ sink checks, and all of them follow the two patterns of \cref{fig:redis_code} and \cref{fig:ip_code}: an object whose array is accessed with a symbolic index, or two fields that one pointer may address. These checks are the evidence that \newdsa{} preserves field sensitivity where \seadsa{} loses it. However, the total reduction is small ($3.9\%$) because few sink checks in these benchmarks depend on either of the two patterns.

With \newdsa{}, the whole pipeline takes $0.8\%$ more time in total (\cref{fig:exp1-table}): $13.9$\,s per benchmark on average against $13.8$\,s with \seadsa{}, and at most $233$\,s in both. The DSA phase takes $0.03$\,s per benchmark on average and $0.6$\,s at most, for both \seadsa{} and \newdsa{}. Both configurations time out on the same $4$ benchmarks, all in the DFA phase, and \cref{fig:exp1-scatter} shows the two configurations close to the diagonal for the others. The largest difference is on IONC, where \newdsa{} takes $8.6\%$ less time in total. A second run of \expone{} gave identical warnings on every benchmark and total running times within $1\%$.

In \exptwo{}, \cref{tab:exp2} reports, for each suite, the number of pointer pairs (all pairs of pointer values within a function, in the functions reachable from \code{main}, that every analysis answers) and the percentage of these pairs that each analysis proves never to alias. A higher percentage means higher precision. LLVM's built-in local alias analysis (BasicAA), which uses no whole-program information, is the baseline, and \svf{} (v3.3) runs its default Andersen analysis and its flow-sensitive variant. \seadsa{} and \newdsa{} run context-insensitive (CI) and context-sensitive (CS), whereas the \svf{} analyses are CI because \svf{}'s CS analysis is demand-driven and aborts on every CURL program. We omit SMALL and $15$ COREUTILS programs ($10$ \svf{} aborts, $4$ query timeouts, one without \code{main}). Under the same sensitivity (CI), the inclusion-based analyses are more precise than the unification-based ones. On IONC, both DSA variants prove at most $13.3\%$ of the pairs because DSA treats a pointer cast from an integer as aliasing any object~\cite{DBLP:conf/pldi/LattnerLA07}, and unification spreads this to every object merged with it~\cite{DBLP:conf/popl/NeculaMW02,DBLP:conf/popl/Steensgaard96}. This cost is independent of the interval cell. These outcomes are expected. The goal of this experiment is to determine whether the interval cell recovers the field sensitivity that DSA loses. On CURL, \newdsa{} proves $1\,704$ pairs that neither \seadsa{} nor \svf{}'s Andersen analysis proves and in which both pointers address one object whose trailing array, as in \cref{fig:redis_code}, is accessed through a pointer with a symbolic offset. \svf{} then makes the whole object field-insensitive and \seadsa{} collapses it, while \newdsa{} keeps its fields separate (\appref{app:field-model} shows the three analyses on this pattern). These results indicate that the loss on this pattern comes from the handling of symbolic offsets and also affects \svf{}, which does not use unification.

\cref{fig:exp2-time} reports the total analysis time per suite, without alias queries, over the programs all four analyses complete. We exclude the $10$ COREUTILS programs where \svf{} aborts. For each analysis, the time is the total running time of the tool, which starts from the LLVM bitcode. \newdsa{} and \seadsa{} take nearly the same time, since the interval cells add at most $11\%$. Andersen takes $3.8\times$, $2.0\times$, and $1.7\times$ the time of \newdsa{} on IONC, CURL, and COREUTILS, and the flow-sensitive analysis takes $8.1\times$, $2.5\times$, and $1.9\times$. No analysis takes more than $5.1$\,s on any benchmark, and \newdsa{} takes at most $1.6$\,s. A second run of \exptwo{} gave identical alias answers on every program and running times within $2\%$.

In \expthree{}, \phasar{} runs the shared policy except for $37$ rules that its configuration cannot express, such as variadic sources and sinks. These rules account for one disagreeing site. On the $155$ programs both tools complete in both runs, they disagree on $92$ sink sites in $46$ programs. The three most common reasons are the following. First, a value computed from a tainted value, such as \code{x + y} with \code{y} tainted, is tainted for us but not for \phasar{}. Second, \phasar{} tracks tainted memory by the pointer that wrote it, whereas we track it by the memory region (DSA cell). When a sink reads the same memory through another pointer, computed by arithmetic or loaded from a structure, only we report it. Third, \phasar{}\footnote{\phasar{} v2604, \code{LLVMAliasSet} over LLVM's \code{CFLAndersAA}, a function analysis.} uses LLVM's alias analysis, which is an intraprocedural analysis. Inside a function, it cannot determine what a pointer parameter points to, so it treats the parameter as a possible alias of global variables such as \code{stderr} and of the other pointer parameters. When the memory a parameter points to is tainted, \phasar{} also taints these globals and parameters, and later calls to \code{fprintf(stderr, ...)} are reported although their messages are clean. Our memory graph covers the whole program, so the parameter points only to the callers' objects. These results indicate that the two tools disagree mainly because they propagate taint under different rules or different memory models. The reason is not that one tool is more precise under the same rules and memory model. The results of two taint analyses are therefore comparable only after their rules and memory models are aligned, which the shared policy alone does not achieve.

\cref{fig:exp3-time} reports the running time of the whole pipeline per benchmark against \phasar{}. Under the same timeout, our analysis completes $193$ of the $197$ benchmarks and \phasar{} $157$ ($158$ in its second run). On SMALL and CURL, both tools complete every program within $2$ seconds, except for one \phasar{} timeout on SMALL. On IONC, our analysis completes more programs and is $37\times$ faster in total on the programs both complete. On the COREUTILS programs both complete, although \phasar{} is faster on $54$, our analysis is faster on the other $28$ and $4.5\times$ faster in total. Outside these cases, \phasar{} times out on $25$ COREUTILS programs and our analysis on $2$ others. Our analysis therefore scales better than \phasar{} on IONC and COREUTILS. A second run of \expthree{} gave the same results on every program \phasar{} completed in both runs, and total running times within $2\%$ on the real-world suites for both tools.

Overall, \newdsa{} preserves field sensitivity where \seadsa{} loses it, at nearly the same running time. However, the reduction in warnings is small on these benchmarks, because few sink checks depend on the two patterns of \cref{fig:redis_code} and \cref{fig:ip_code}. The first pattern, an array accessed with a symbolic index, also affects \svf{}, which does not use unification: \svf{} loses field sensitivity when the array is accessed through a pointer. Against \phasar{}, our taint analysis completes more benchmarks, and the two tools disagree mainly because their propagation rules and memory models differ.
 \section{Related Work}\label{sec:related}

Many static taint analyzers formulate taint propagation as a data-flow problem but differ in flow, field, and context sensitivity and in their heap abstraction.

Precision in memory-aware taint tracking is limited by the accuracy of the underlying points-to analysis (PTA). A common strategy is to perform PTA to model memory, followed by a data-flow analysis (DFA) for taint propagation.
We use a unification-based pointer analysis~\cite{DBLP:conf/popl/Steensgaard96} called Data Structure Analysis (DSA)~\cite{DBLP:conf/pldi/LattnerA05,DBLP:conf/pldi/LattnerLA07} for memory-effects inference. Other approaches, such as \comptaint~\cite{DBLP:conf/sigsoft/BanerjeeCEFHLLP23} and \ptaint~\cite{DBLP:journals/pacmpl/GrechS17}, also combine the two analyses with different designs.
\comptaint{} runs a heap analysis to generate per-function points-to summaries, similar to DSA, and extends them directly with taint data-flow facts. This yields reusable summaries that can be instantiated at different call sites to recover context sensitivity, avoiding whole-program reasoning for scalability. \ptaint{} also integrates DFA but tracks taint flows inside a pointer analysis and updates taint facts as it discovers points-to information. This unified design allows points-to information to inform taint propagation immediately.

Our approach refines DSA through partial node collapse, whereas standard DSA collapses a node, merging all its cells, after any incompatible unification. Our approach uses interval offsets to localize cell merges: on a unification conflict, we merge only the affected cells, and the remaining cells keep their precision. This directly improves taint-tracking precision. Other analyses also improve DSA field sensitivity. \seadsa~\cite{DBLP:conf/sas/GurfinkelN17} formalizes unification for array-representative nodes. \teadsa~\cite{DBLP:conf/fmcad/KuderskiNG19} leverages type information to reduce type-incompatible unifications. Our approach is closest to \dsabin~\cite{dsabin}, which uses constant-set offset abstractions to prevent full node collapse. However, where \dsabin{} collapses a node when two of its cells are unified, we merge only the cells between them. Unlike \dsabin{}, which focuses on binary analysis, we analyze LLVM IR, the original target of DSA. Abstract domains for array segmentation~\cite{DBLP:conf/popl/CousotCL11} divide an array into consecutive segments and summarize the elements of each segment separately. \newdsa{} instead summarizes in one interval cell the elements that a symbolic index may reach, and keeps the fields outside this interval separate.

Our DFA treats taint facts as lattice elements to identify tainted variables or memory locations. Other taint-tracking strategies have also been proposed. A common framework is \emph{Interprocedural Finite Distributive Subset} (IFDS)~\cite{DBLP:conf/popl/RepsHS95}, which constructs an \emph{Exploded Super-Graph} with instruction-fact pairs as nodes and data-flow edges. IFDS reduces data-flow analysis to graph reachability. \emph{Interprocedural Distributive Environment} (IDE)~\cite{DBLP:journals/tcs/SagivRH96} generalizes IFDS. Practical implementations include \phasar~\cite{DBLP:conf/tacas/SchubertHB19}, which applies IFDS and IDE to analyze LLVM IR programs and builds a taint analysis over a taint-fact domain, and \flowdroid~\cite{DBLP:conf/pldi/ArztRFBBKTOM14}, which uses IFDS for taint analysis on Android apps.

Our DFA follows abstract interpretation~\cite{DBLP:conf/popl/CousotC77} to approximate taint flow. It relates to the abstract semantic dependency framework introduced by Cousot~\cite{DBLP:conf/sas/Cousot19a}, where taint analysis is formulated as a sound abstraction for tracking value dependencies. At each program point, the analysis over-approximates the set of variables whose values may depend on tainted inputs. \mopsa~\cite{DBLP:conf/vmcai/ParoliniM24} also introduces taint analysis to verify that attacker-controlled inputs cannot reach and trigger critical runtime errors (i.e., nonexploitability). Both analyses use numerical domains for precision but differ in scope and granularity. \mopsa{} focuses on inputs that can trigger runtime errors, whereas our work identifies data leaks to sinks such as diagnostic logging. Unlike \mopsa{}, which does not track taint at the level of individual memory locations (field-insensitive), our work uses \newdsa{} to provide the field sensitivity required by this objective.

\section{Conclusion}

In this work, we present a taint analysis for LLVM IR that follows user-specified taint sources, sinks, and propagation rules. The analysis combines Data Structure Analysis (DSA) for memory modeling with data-flow analysis (DFA) for taint propagation. We introduce \newdsa{}, a DSA refinement that improves field sensitivity, and build our DFA on top of an existing abstract-interpretation-based value analysis to prune infeasible taint facts. Our results show that this design provides a practical precision-cost trade-off for memory-aware taint analysis.
 
\begin{credits}
\subsubsection{Data Availability Statement.}
The artifact with the tools, benchmarks, and data used for the experimental evaluation in this paper is available at~\cite{artifact}.
\end{credits}
\bibliographystyle{splncs04}

\appendix
\section{Soundness Proofs}
\label{proof:soundness}

\subsection{Soundness of the Pointer Analysis}
\label{proof:dsa-soundness}

We first show that unification preserves a simulation (\cref{lemma:unify-sim}) and use it to prove \cref{thm:dsa-soundness} for primitive statements, case by case. We then show that the iteration with widening stops with a graph that simulates every reachable state (\cref{prop:dsa-iteration}), and extend \cref{thm:dsa-soundness} to the other statements. All proofs follow the semantics of \cref{sec:concrete_lan_sem}. Given a state $\state$ and a graph $\graph$, we ignore uninitialized pointer variables when checking $\state \models_\mu \graph$. That is, we consider only pointer variables in $\varptrset \cap \dom(\cenv)$. We also assume that a pointer variable holds only addresses and an integer variable only integers, so a load into a pointer variable reads an address, and that \code{input} is used on integer variables only.

\begin{lemma}[Unification preserves the simulation]
\label{lemma:unify-sim}
Let $\state \models_\mu \graph$, and let a call of $\unifyCells$ on two cells of $\graph$ turn $\graph$ into $\graph'$. For each cell $\cell$ of $\graph$, let $\pi(\cell)$ denote the cell of $\graph'$ representing it after unification. For each original node $\node$, its cells move together into one final node with a common cumulative shift $\delta_\node \geq 0$, and $\pi(\cell)$ covers the offsets of $\cell$ shifted by $\delta_\node$. Then $\state \models_{\pi \circ \mu} \graph'$. The same holds for $\mergeOrNew$, also when the merged cell gets a larger interval.
\end{lemma}
\begin{proof}
Let the call be $\unifyCells(\cell_1, \cell_2)$ with $\omin{\offsetfunc{\cell_1}} \leq \omin{\offsetfunc{\cell_2}}$. It only changes the graph, not $\state$. We follow the algorithm step by step.

\textbf{Same node ($\unifyCellsSame$).} We merge the cells that overlap the join of the intervals of $\cell_1$ and $\cell_2$ into one cell $\cell_m$. A location that $\mu$ mapped to one of these cells is now represented by $\cell_m$. Since the interval of $\cell_m$ covers the intervals of all merged cells, $\cell_m$ still covers the offset of the location, and this merge changes no object anchor.

\textbf{Different nodes ($\unifyCellsOther$).} We merge every cell of $\node_1$ into $\node_2$ at the relative offset $\delta = \omin{\offsetfunc{\cell_2}} - \omin{\offsetfunc{\cell_1}} \geq 0$, and then remove $\node_1$. A location that $\mu$ mapped to a cell of $\node_1$ is now represented by a cell of $\node_2$ whose interval covers the shifted offset; that is, an object placed at offset $x$ in $\node_1$ is now placed at $x + \delta$ in $\node_2$. In particular, the shifted $\cell_1$ starts where $\cell_2$ starts, so $\cell_1$ is merged with $\cell_2$.

\textbf{Edges ($\redirectEdges$).} In both cases, we redirect each merged cell $\cell_i$ to its new cell $\cell_m$. First, every pointer or edge that points to $\cell_i$ now points to $\cell_m$, which now represents the addresses previously represented by $\cell_i$. Second, the outgoing edge of $\cell_i$ moves to $\cell_m$, so a location of $\cell_i$ that stores an address still leads to the cell of that address. If $\cell_i$ and $\cell_m$ both have an outgoing edge, we recursively unify their targets so that the merged cell has a single outgoing edge, as required by C.3. We follow node and cell forwarding throughout the recursion. The recursion ends because every nontrivial unification merges two current nodes or at least two current cells on the same node, and unifying a cell with itself is a no-op.

Let $\pi(\cell)$ denote the cell representing $\cell$ after unification; in particular, $\pi(\cell_1) = \pi(\cell_2)$. As shown above, each location that $\mu$ mapped to $\cell$ is represented by $\pi(\cell)$, so $\state \models_{\pi \circ \mu} \graph'$. For $\mergeOrNew$, if no cell overlaps the requested interval ($S = \emptyset$), creating a new cell leaves the existing simulation unchanged. Otherwise, we join the requested interval with the intervals of all overlapping cells. Since the original cell intervals are disjoint, this join overlaps no remaining cell, so the merge preserves C.1. The edges of merged cells are redirected as shown above, which preserves C.3.
\end{proof}

\begin{proof}[Proof of \cref{thm:dsa-soundness} for primitive statements]
Take any state $\state \in \concretegraph(\graph)$ with a simulation $\mu$ such that $\state \models_\mu \graph$. Let $\state' := \concretetran(\state)$ be the state after the statement and $\graph' := \dsatran(\graph)$ the graph after it. We construct a simulation $\mu'$ such that $\state' \models_{\mu'} \graph'$. Every transfer function in \cref{fig:dsa-trans} ends with a unification, which preserves the simulation by \cref{lemma:unify-sim}, and $\lookUp$ only adds nodes and cells that no location is mapped to; so for each statement, we only check the points-to relation that the statement adds. The taint component plays no role here.

\textbf{Allocation $\pointerp = \makeref(\inti)$.} We assume that allocation creates a fresh, uninitialized object, so the heap initially has no entries for it. The concrete semantics assigns the object a fresh base $\baddr$ and lets $\pointerp$ refer to $(\baddr, 0)$. Following the rule, we create a new node $\freshnode$ with the cell $(\freshnode, [0,0])$ and map $(\baddr, 0)$ to it, which places the object at offset $0$. We then unify the cell of $\pointerp$ with $(\freshnode, [0,0])$. Hence, with $\mu' := \pi \circ \update{\mu}{(\baddr, 0) \mapsto (\freshnode, [0,0])}$, $\pointerp$ points to the cell of $(\baddr, 0)$.

\textbf{Pointer arithmetic $\pointerp = \pointerq + \of$.} Let $\pointerq$ hold $(\baddr,\of_\pointerq)$, which $\mu$ maps to $(\node,\absof_\pointerq)$ with the object at offset $x$ of $\node$. Then $x+\of_\pointerq \in \offsetgamma(\absof_\pointerq)$, so the new address $(\baddr,\of_\pointerq+\of)$ lies at offset $x+\of_\pointerq+\of \in \offsetgamma(\absof)$ of $\node$, where $\absof = \absof_\pointerq \offsetadd [\of,\of]$. The rule builds the cell $\cell_m$ of $\node$ that covers $\absof$ ($\mergeOrNew$, cell map $\pi_1$) and unifies the cell of $\pointerp$ with it (cell map $\pi_2$), so by \cref{lemma:unify-sim}
\[ \mu' := \pi_2 \circ \update{(\pi_1\circ\mu)}{(\baddr,\of_\pointerq+\of) \mapsto \cell_m}. \]
If the address was already mapped, its cell covers the same offset of $\node$ and is $\cell_m$ by C.1.

\textbf{Array pointer arithmetic $\pointerp = \pointerq + (\inti \times \elementsize)$.} The concrete semantics lets $\pointerp$ refer to $(\baddr, \of_\pointerq + \cenv(\inti) * \elementsize)$, where $\cenv(\inti) * \elementsize \geq 0$ by the restriction of \cref{sec:concrete_lan_sem}. Since $\absof = \absof_\pointerq \offsetadd [0, +\infty]$ covers every such offset, the argument is the same as for $\pointerp = \pointerq + \of$.

\textbf{Load $\varx = *\pointerp$, where $\varx$ is a pointer.} Let $\addrval' = \heap(\cenv(\pointerp))$ be the address loaded into $\varx$. Since $\state \models_\mu \graph$, $\mu$ maps $\addrval'$ to the target of the cell of $\pointerp$ (denoted as $\cell_d$). Following the rule, we unify $\cell_d$ with the cell of $\varx$. Hence, with $\mu' := \pi \circ \mu$, $\varx$ points to the cell of $\addrval'$.

\textbf{Store $*\pointerp = \varx$, where $\varx$ is a pointer.} The concrete semantics stores the address held by $\varx$ (denoted as $\addrval_\varx$) at the address held by $\pointerp$. Following the rule, we find the target $\cell_d$ of the cell of $\pointerp$, which is a fresh cell if the cell of $\pointerp$ has no outgoing edge, and unify $\cell_d$ with the cell of $\varx$, which $\mu$ maps $\addrval_\varx$ to. Hence, with $\mu' := \pi \circ \mu$, the edge from the cell of $\pointerp$ leads to the cell of $\addrval_\varx$, as the new heap entry requires.

\textbf{Other statements.} The statements without a rule in \cref{fig:dsa-trans} leave $\graph$ unchanged: integer assignments, input of an integer, sanitizers, output, and loads and stores of integers. They only change integer values, or the value stored at the address held by $\pointerp$, which $\mu$ already maps. We have $\mu' := \mu$.

In every case, $\state' \models_{\mu'} \graph'$, so \cref{thm:dsa-soundness} holds for primitive statements.
\end{proof}

\textbf{The iteration.} We write $\graph \sqsubseteq \graph'$ if a finite sequence of unification steps turns $\graph$ into $\graph'$:
\[
\graph \sqsubseteq \graph' \iff \graph' = u_k(\cdots u_1(\graph)\cdots) \text{ for some } k \geq 0,
\]
where each $u_i$ is a call of $\unifyCells$ on two cells of its argument, or a call of $\mergeOrNew$ with an interval that overlaps a cell of its node, with $S$ the cells of that node that the interval overlaps. Unification never splits a node or a cell, and intervals are only joined, or shifted with their node when it is merged into another, so each $u_i$ merges nodes or cells, increases the interval of a cell, or changes nothing. Hence $\graph \sqsubseteq \graph'$ means that $\graph'$ is $\graph$ with some nodes and cells merged and some intervals increased, and two facts follow. First, $\concretegraph(\graph) \subseteq \concretegraph(\graph')$, since each $u_i$ preserves simulations (\cref{lemma:unify-sim}). Second, $\graph \sqsubseteq \graph' \sqsubseteq \graph$ implies $\graph' = \graph$: a step that changes a graph lowers its number of nodes and cells or, keeping this number, increases an interval, so no sequence of steps returns to a graph it has changed.

Let $F^\sharp$ apply the transfer function of every primitive statement once, in a fixed syntactic order. Let $F^\sharp_\nabla$ be $F^\sharp$ with the join replaced by the widening~\cite{CousotIntervals} where pointer arithmetic grows the cell of $\pointerp$: if the interval $[a,b]$ of this cell does not contain the interval $\absof$ that the rule requires, the cell gets $[a,b] \mathbin{\nabla} ([a,b] \offsetjoin \absof)$ instead of $[a,b] \offsetjoin \absof$, and $\mergeOrNew$ merges it with every cell that this interval overlaps, as C.1 requires. Both $[a,b]$ and $\absof$ are intervals of the node of $\pointerq$: the first application of the rule put the cells of $\pointerp$ and $\pointerq$ in one node, and unification moves the cells of a node together. As usual, we delay the widening by one step: a singleton $[a,a]$ still grows by join, so that a single field offset is not lost on its first growth. The joins inside unification stay joins. The analysis applies $F^\sharp$ once and then $F^\sharp_\nabla$ until $F^\sharp$ no longer changes the graph:
\[
\graph_0 = F^\sharp(\graph_\emptyset), \qquad
\graph_{k+1} =
\begin{cases}
\graph_k & \text{if } F^\sharp(\graph_k) \sqsubseteq \graph_k \quad\text{(stop)},\\
F^\sharp_\nabla(\graph_k) & \text{otherwise,}
\end{cases}
\]
where $\graph_\emptyset$ is the empty graph. In later iterations, an allocation unifies its fresh node with the existing node of $\pointerp$; this changes at most the name of that node, not its cells, so we treat graphs that differ only in node names as equal and identify their cells.

From $\graph_0$ on, the transfer functions no longer add nodes or cells. Every pointer variable and every dereference already has its cell. In pointer arithmetic, the cell of $\pointerp$ already overlaps the required interval, because both contain the interval required in the previous iteration, shifted with their node. Hence every transfer function is a sequence of steps $u_i$, and the iterates form an increasing chain $\graph_0 \sqsubseteq \graph_1 \sqsubseteq \cdots$. Moreover, $F^\sharp$ and $F^\sharp_\nabla$ differ only where the cell of a pointer must grow. So a graph that $F^\sharp_\nabla$ leaves unchanged is also left unchanged by $F^\sharp$.

\begin{lemma}[Widening]
\label{lemma:widening}
For nonempty intervals of non-negative offsets, the widening $\nabla$ of \cref{sec:dsa_o} satisfies: (i) $[a,b] \offsetjoin [c,d] \offsetleq [a,b] \mathbin{\nabla} [c,d]$; (ii) for every sequence of intervals $(y_i)_{i \in \mathbb{N}}$, the sequence $x_0 = y_0$, $x_{i+1} = x_i \mathbin{\nabla} y_{i+1}$ is eventually stable.
\end{lemma}
\begin{proof}
The lower bound of $[a,b] \mathbin{\nabla} [c,d]$ is $a$ if $a \leq c$ and $0$ otherwise, so it is at most $\min(a,c)$; symmetrically, the upper bound is at least $\max(b,d)$. This gives (i). A lower bound $0$ never changes again, nor does an upper bound $+\infty$, so each bound of $x_i$ changes at most once. This gives (ii).
\end{proof}

\begin{proposition}[Iteration with widening]
\label{prop:dsa-iteration}
For a finite program, the sequence $(\graph_k)$ is eventually stable. Its limit $\graph$ is a fixpoint of every transfer function of \cref{fig:dsa-trans} and simulates every reachable state.
\end{proposition}
\begin{proof}
\textbf{Termination.} The iterates form an increasing chain, so we only need to show that it cannot grow forever. The graph grows in two ways. First, nodes or cells are merged by unification. Each merge reduces their number, so only finitely many merges can happen. Second, once merges stop, the graph changes only where pointer arithmetic grows a cell. A cell grows by join at most once, while it is a singleton, and by widening afterwards, so by \cref{lemma:widening}(ii) each of the finitely many intervals stabilizes. Thus $F^\sharp_\nabla$ eventually leaves the graph unchanged. Then $F^\sharp$ does too, and the stop case applies.

\textbf{Soundness.} Let $\graph$ be the limit, and let $\graph^{(0)} = \graph, \graph^{(1)}, \ldots, \graph^{(m)} = F^\sharp(\graph)$ be the graphs after each transfer function of $F^\sharp$. Each $\graph^{(i+1)}$ follows from $\graph^{(i)}$ by steps of $\sqsubseteq$, and the stop case gives $F^\sharp(\graph) \sqsubseteq \graph$:
\[
\graph = \graph^{(0)} \sqsubseteq \graph^{(1)} \sqsubseteq \cdots \sqsubseteq \graph^{(m)} \sqsubseteq \graph .
\]
By transitivity and the second fact above, every $\graph^{(i)}$ equals $\graph$, so no transfer function changes $\graph$, and by \cref{thm:dsa-soundness} every primitive statement maps $\concretegraph(\graph)$ into itself. The initial state, with an empty environment and an empty heap, is in $\concretegraph(\graph)$ with the empty $\mu$. By induction on the program, guards, sequences and branches preserve containment in $\concretegraph(\graph)$; for a loop, $\concretegraph(\graph)$ is a post-fixpoint of its functional in \cref{fig:trans} and so contains the least fixpoint by Tarski's theorem~\cite{Tarski55}. Hence $\graph$ simulates every reachable state.
\end{proof}

\textbf{Non-primitive statements.} For such a statement $\statements$ and a graph $\graph'$, $\dsatran(\graph')$ is the limit $\graph$ of the iteration above, where the iteration starts from $\graph'$ instead of $\graph_\emptyset$ and $F^\sharp$ applies the transfer functions of the primitive statements in $\statements$. As before, the iteration stops, and these transfer functions do not change $\graph$. They only unify cells or add cells that no location is mapped to, so $\concretegraph(\graph') \subseteq \concretegraph(\graph)$ by \cref{lemma:unify-sim}. The induction above, applied to $\statements$, shows that $\concretetranon{\statements}$ maps $\concretegraph(\graph)$ into itself, so $\concretetranon{\statements}(\concretegraph(\graph')) \subseteq \concretetranon{\statements}(\concretegraph(\graph)) \subseteq \concretegraph(\graph)$.

\subsection{Soundness of the Taint Analysis}
\label{proof:taint-soundness}

\begin{proof}[Proof of \cref{thm:dfa-soundness}]
The taint analysis runs on the final DSA graph $\graph$, which the transfer functions of \cref{fig:dsa-trans} no longer change. For this $\graph$, the proof of \cref{thm:dsa-soundness} gives, for every primitive statement, a simulation $\mu'$ of the next memory state that agrees with $\mu$ on previously mapped addresses and extends it when allocation or pointer arithmetic produces an unmapped address.

Take an abstract state $\absstate = (\absmemstate, \abstaintstate)$ and a concrete state $(\cenv, \heap, \taintMap) \in \concretetaint(\absstate)$ with its simulation $\mu$. Let $(\cenv', \heap', \taintMap')$ be the state after the statement, that is, $\taintctran((\cenv, \heap, \taintMap))$. We show that $(\cenv', \heap', \taintMap') \in \concretetaint(\dfatran(\absstate))$. For the memory part, $(\cenv', \heap') \in \concretemem(\memtran(\absmemstate))$ because the memory transfer function is sound, and $(\cenv', \heap') \models_{\mu'} \graph$ as above. It remains to show that every location in $\taintMap'$ is represented in the new taint set $\abstaintstate_1$: a tainted variable is in $\abstaintstate_1$, and a tainted address $\addrval$ has its cell $\mu'(\addrval)$ in $\abstaintstate_1$.

\textbf{Primitive statements.} For \code{input}, both rules add the variable to the taint set, and for \code{sanitize}, both remove it. For an assignment, pointer arithmetic and allocation, both rules use the same sources and targets, which are variables. If a source is tainted concretely, it is also in $\abstaintstate$, so the abstract rule marks the target as tainted. The abstract rule clears the target variable's previous taint and marks it as tainted again if any source is in the input taint set. Thus a target that becomes clean concretely may remain tainted abstractly, which is sound.

For a load $\varx = *\pointerp$, the concrete rule taints $\varx$ if the address $\addrval_\pointerp = \cenv(\pointerp)$ is tainted. Then its cell $\mu(\addrval_\pointerp)$ is in $\abstaintstate$, and this cell is the cell $\cell = \graph.\envmap(\pointerp)$ that $\pointerp$ points to, because $\state \models_\mu \graph$. So the abstract rule $\abspropshort{\{\cell\}}{\{\varx\}}$ also taints $\varx$. For a store $*\pointerp = \varx$, the concrete rule taints $\addrval_\pointerp$ if $\varx$ is tainted. Then $\varx \in \abstaintstate$, so the abstract rule $\abspropshort{\{\varx\}}{\{\cell\}}$ adds $\cell = \mu'(\addrval_\pointerp)$. If $\varx$ is concretely clean, the concrete rule clears the taint at $\addrval_\pointerp$. The abstract rule never clears taint for memory cells and may also add $\cell$ if $\varx$ is abstractly tainted. Both possibilities soundly over-approximate the concrete result. In every case, the other tainted addresses keep their cells, because $\mu'$ agrees with $\mu$ on them. Finally, $\koutput(\varx)$ changes neither state; the analysis reports a warning there if $\varx \in \abstaintstate$.

\textbf{Guards and composition.} Evaluating a guard does not change the taint set, and the memory domain filters $\absmemstate$ soundly. If the filtered memory state is $\membot$, the reduction $\mathcal{R}$ empties the taint set, which is sound because $\concretemem(\membot)$ is empty. The abstract states are ordered componentwise, by the order of the memory domain and by $\subseteq$ on taint sets, and joined by $\memjoin$ and $\cup$; $\concretetaint$ is monotone in both components. So a sequence of statements composes the argument above, and the join at the end of an \code{if} covers the states of both branches. For a loop, iteration with widening computes a post-fixpoint of the abstract loop functional, which joins the abstract entry state with the result of applying the guarded body to the current abstract state. Its concretization contains the entry states and is preserved by the guarded body, so it is an inductive invariant containing the least fixpoint in \cref{fig:trans}. Applying the exit guard then gives a sound approximation of the states after the loop. The taint component needs no widening, since the set of abstract locations is finite.

Hence $\taintctran(\concretetaint(\absstate)) \subseteq \concretetaint(\dfatran(\absstate))$, so \cref{thm:dfa-soundness} holds. In particular, since a tainted variable is always in $\abstaintstate$, the absence of a warning at $\koutput(\varx)$ means that $\varx$ is untainted in every execution that reaches it.
\end{proof}
\section{Field Sensitivity of \svf{}, \seadsa{} and \newdsa{} on One Object}
\label{app:field-model}

\cref{fig:app-field-code} is the function of \cref{fig:redis_code} with the flexible array filled through a pointer, \code{int *p = b->buf; p[i] = s[i]}: a store through a pointer at an index \code{i} that is not a constant. \cref{fig:app-field-graphs} shows the object allocated by \code{malloc} in three of the pointer analyses of \exptwo{}, cut from the graphs the tools dump: the pointers of the function into the object and the cell or points-to target each one reaches. Like the benchmarks of \cref{sec:eval}, the function is compiled for a 64-bit target, where \code{size_t} takes $8$ bytes, so the offsets differ from those in \cref{fig:flow_redis}.

\begin{figure}[t]
\centering
\begin{lstlisting}[basicstyle=\footnotesize, language=C, style=mystyle, xleftmargin=11.81pt]
void fRepBuf(const int *s, size_t len) {
  repl_block_t *b = malloc(sizeof(repl_block_t) + len * 4);
  b->id = block_id++;  b->size = len;  b->used = 0;
  size_t copy = len < b->size - b->used ? len : b->size - b->used;
  int *p = b->buf;                  // the array reached through a pointer
  for (size_t i = 0; i < copy; ++i)
    p[i] = s[i];                    // index i is not a constant
  b->used += copy;
  log("block id=%d used=%zu size=%zu", b->id, b->used, b->size);
}
\end{lstlisting}
\caption{The function of \cref{fig:redis_code} with the flexible array filled through a pointer.}
\label{fig:app-field-code}
\end{figure}

\begin{figure}[t]
\centering
\begin{minipage}[c]{0.76\linewidth}\centering\includegraphics[width=\linewidth]{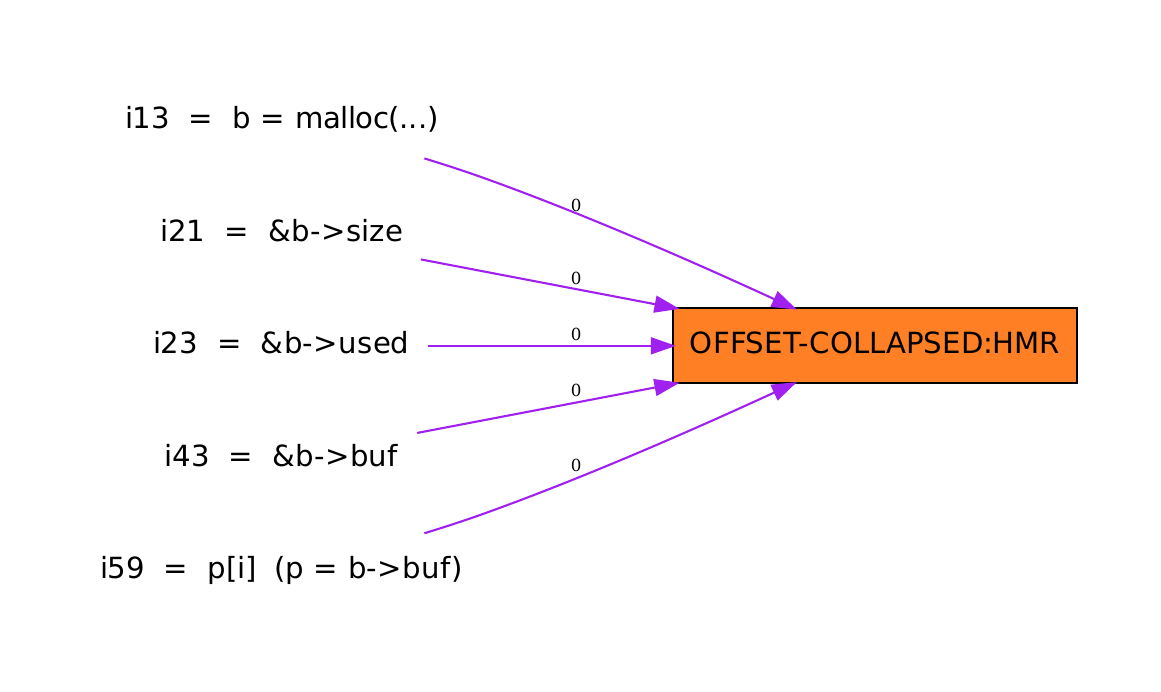}\subcaption{\seadsa{}: one collapsed cell, every pointer at offset $0$.}\end{minipage}\\[4pt]
\begin{minipage}[c]{0.76\linewidth}\centering\includegraphics[width=\linewidth]{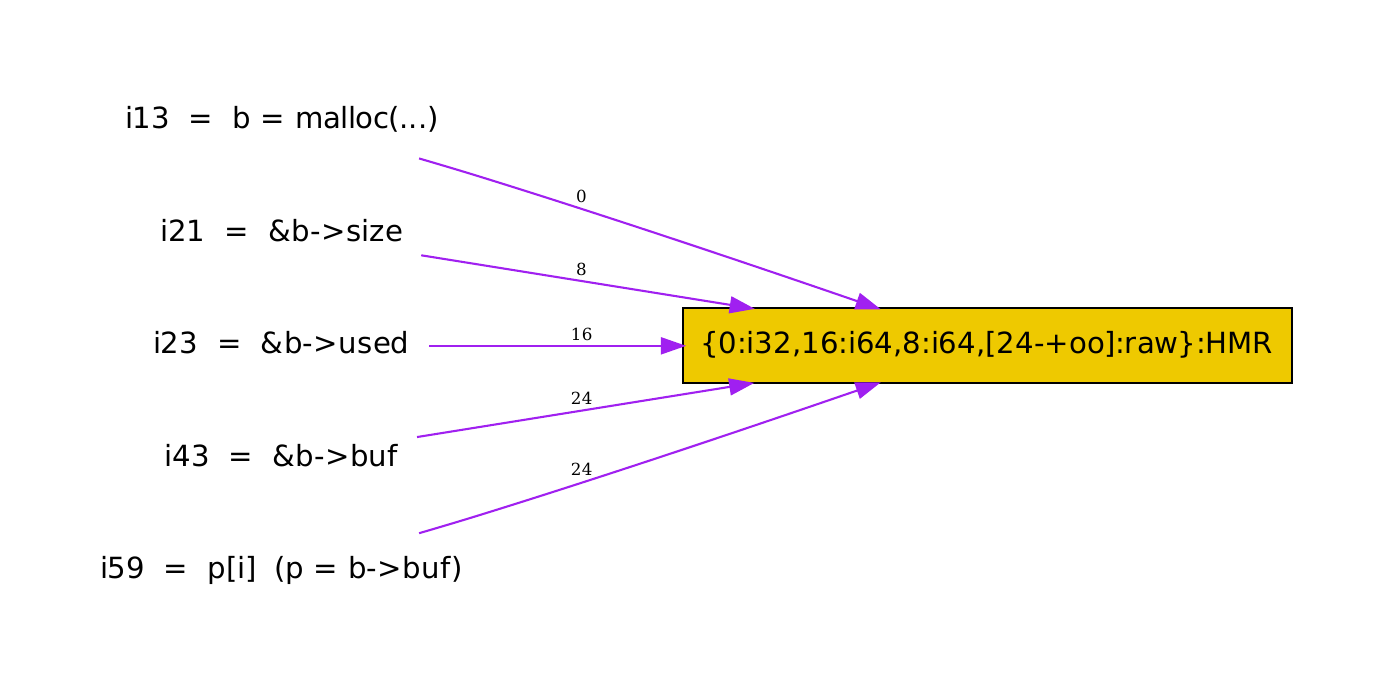}\subcaption{\newdsa{}: cells at $0$, $8$, $16$ and the interval cell $[24,+\infty]$.}\end{minipage}\\[4pt]
\begin{minipage}[c]{0.76\linewidth}\centering\includegraphics[width=\linewidth]{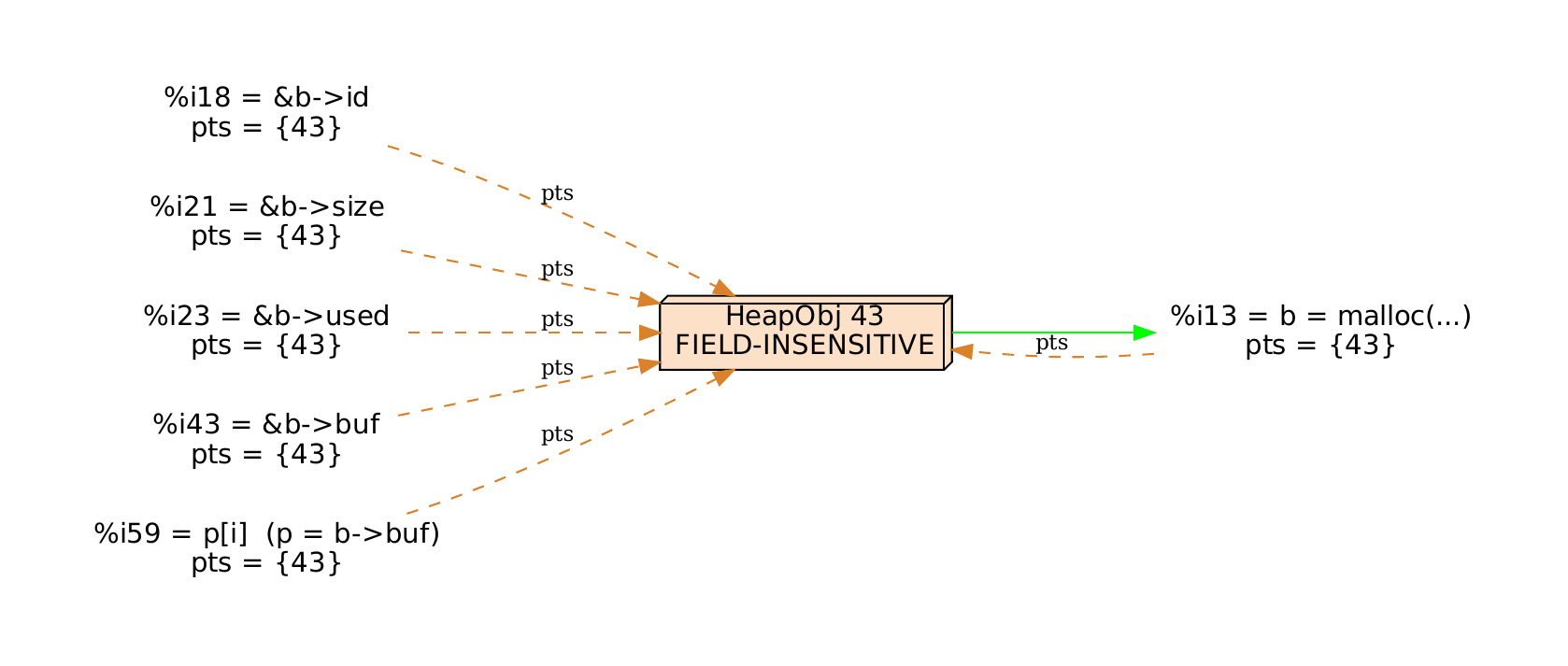}\subcaption{\svf{} Andersen: one field-insensitive object, points-to set $\{43\}$ for every pointer.}\end{minipage}
\caption{The object of \cref{fig:app-field-code} in the three analyses. In (a) and (b), edge labels are byte offsets, \texttt{b} also stands for \texttt{\&b->id}, which the two tools do not print separately, and \texttt{HMR} marks a heap node that is modified and read. In (c), dashed edges are the printed points-to sets, and the solid edge is the address edge of \svf{} from the heap object to the pointer that \texttt{malloc} returns.}
\label{fig:app-field-graphs}
\end{figure}

\textbf{SeaDSA.}
Unification places every pointer of the function on the node of \code{b}. The store through \code{p} adds an access at a symbolic offset to the array cell; DSA has no cell for a symbolic offset and collapses the node, so \code{\&b->id}, \code{\&b->size}, \code{\&b->used} and \code{p[i]} all reach one \code{OFFSET-COLLAPSED} cell at offset~$0$ (\cref{fig:app-field-graphs}a). The fields are lost.

\textbf{\newdsa{}.}
The same unification, but the symbolic offset becomes the interval cell $[24, +\infty]$, from the start of \code{buf} to the end of the object, and the cells at $0$, $8$ and $16$ stay: \code{\&b->id}, \code{\&b->size} and \code{\&b->used} reach their own cells and \code{p[i]} the interval cell (\cref{fig:app-field-graphs}b). The fields are kept; the array elements are one cell, as in every analysis.

\textbf{SVF Andersen.}
\svf{} maps every field access to a field index of the flattened structure and creates one field object per index. The store through \code{p} adds a non-constant offset to a plain \code{int} pointer and cannot be mapped to a field; \svf{} records it as a \emph{variant} access, and while solving marks every object \code{p} may point to as field-insensitive and folds the field objects into the base. Every pointer into the object, \code{\&b->id} and \code{\&b->size} as much as \code{p[i]}, then has the points-to set $\{43\}$ (\cref{fig:app-field-graphs}c). The fields are lost. With the direct index of \cref{fig:redis_code}, \code{b->buf[i]}, \svf{} drops the array index, the access stays in the field of \code{buf}, and the four field objects survive; \seadsa{} and \newdsa{} behave as above in both cases.

The loss on this object therefore comes from how an analysis handles a symbolic offset: \seadsa{} loses the fields whenever the offset is symbolic, \svf{} when the symbolic offset is applied to a pointer, and \newdsa{} keeps them in both cases by bounding the offset with an interval instead of abandoning the cells. On CURL, such objects account for the $1\,704$ pairs of \cref{sec:eval}, $1\,555$ of them in the driver of \code{Curl\_he2ai}.
 
\end{document}